\documentclass{aa}
\usepackage[varg]{txfonts}
\usepackage{natbib}
\bibpunct{(}{)}{;}{a}{}{,} 

\begin{document}

\title{Compositional study of asteroids in the Erigone collisional family using visible spectroscopy at the 10.4m GTC}
\titlerunning{Visible spectroscopy of Erigone collisional family}

\author{David Morate\inst{1,2}
     \and Julia de Le\'on\inst{1,2}
     \and M\'ario De Pr\'a\inst{3}
     \and Javier Licandro\inst{1,2}
     \and Antonio Cabrera-Lavers\inst{1,4}
     \and Humberto Campins\inst{5}
     \and Noem\'i Pinilla-Alonso\inst{6}
     \and V\'ictor Al\'i-Lagoa\inst{7}}

\institute{Instituto de Astrof\'isica de Canarias (IAC), C/V\'ia L\'actea s/n, 38205 La Laguna, Tenerife, Spain
  \and Departamento de Astrofísica, Universidad de La Laguna, 38205 La Laguna, Tenerife, Spain
  \and Observat\'orio Nacional, Coordenaç\~ao de Astronomia e Astrofísica, 20921-400 Rio de Janeiro, Brazil
  \and GTC Project Office, 38205 La Laguna, Tenerife, Spain
  \and Physics Department, University of Central Florida, P.O. Box 162385, Orlando, FL 32816-2385, USA
  \and Department of Earth and Planetary Sciences, University of Tennessee, Knoxville, 37996 TN, USA
  \and Laboratoire Lagrange, OCA, Boulevard de l'Observatoire, B.P. 4229 06304 Nice Cedex 04 - France}

\abstract{Two primitive near-Earth asteroids, (101955) Bennu and (162173) Ryugu, will be visited by a spacecraft with the aim of returning samples back to Earth. Since these objects are believed to originate in the inner main belt primitive collisional families (Erigone, Polana, Clarissa and Sulamitis) or in the background of asteroids outside these families, the characterization of these primitive populations will enhance the scientific return of the missions. The main goal of this work is to shed light on the composition of the Erigone collisional family by means of visible spectroscopy. Asteroid (163) Erigone has been classified as a primitive object \citep{bus99phd,busbinzel02}, and we expect the members of this family to be consistent with the spectral type of the parent body. We have obtained visible spectra (0.5-0.9 $\mu$m) for 101 members of the Erigone family, using the OSIRIS instrument at the 10.4m Gran Telescopio Canarias. We found that $87\%$ of the objects have typically primitive visible spectra consistent with that of (163) Erigone. In addition, we found that a significant fraction of these objects ($\sim50\%$) present evidence of aqueous alteration.}

\keywords{minor planets, asteroids: general - methods: data analysis - techniques: spectroscopic}
\maketitle 


\section{Introduction}
\label{introduction}

Primitive asteroids are considered to be composed of the most pristine materials in the Solar System, being remnants of the processes which followed the condensation of the proto-planetary nebula and the formation of the planets. The materials on these objects have been altered over time due to different processes, such as space weathering and aqueous alteration \citep{fornasier14}. Aqueous alteration acts mainly on primitive asteroids (C, B, and low albedo X-types, according to the \cite{busdemeo2009} classification scheme), producing a low-temperature (< 320K) chemical alteration of the materials due to the presence of liquid water. This water acts as a solvent and generates hydrated materials like phyllosilicates, sulfates, oxides, carbonates, and hydroxides. The presence of hydrated materials thus implies that liquid water was present in the primordial asteroids, produced by the melting of water ice by a heating source \citep{fornasier14}. The most unambiguous indicator of hydration is the 3 $\mu$m hydration band observed in infrared photometry and spectroscopy of many primitive asteroids. This feature is correlated with the 0.7 $\mu$m Fe$^{2+}$ $\rightarrow$ Fe$^{+3}$ oxidized iron absorption band observed in the visible spectra of these asteroids \citep{vilas94,howell11,rivkin12}. 

Apart from this aqueous alteration, primitive asteroids have undergone minimal geological or thermal evolution, experiencing, on the contrary, an intense collisional evolution that affected their shape, size, and surface composition. Therefore, studying the products of those collisional events and their mineralogical compositions, will shed light on the evolutionary history of the Solar System.

Asteroid collisional families are groups of asteroids sharing very similar orbital properties \citep{hirayama18}, thought to be the direct result of energetic collisional events. Spectroscopic observations provided the first confirmation of the collisional origin of a family: the Vesta family \citep{binzelxu93}. The spectroscopic study of collisional families have steadily increased since then, focusing on the characterization of their mineralogy. Additionally, considering that near-Earth asteroids (NEAs) come primarily from the main asteroid belt \citep{bottke02}, collisional families are good sources of NEAs, as they generate plenty of small fragments during their formation. Families located close to particular resonances in the belt can easily send these fragments to the near-Earth space. In this sense, the inner asteroid belt (the region located between the $\nu_6$ resonance, near 2.15 AU, and the 3:1 mean-motion resonance with Jupiter, at 2.5 AU) is considered as a primary source of near Earth asteroids \citep{bottke02}.

NASA OSIRIS-REx \citep{lauretta10} and JAXA\footnote{Japan Aerospace Exploration Agency} Hayabusa 2 \citep{tsuda13} sample-return missions have targeted two near Earth asteroids: (101955) Bennu and (162173) Ryugu, respectively. These are primitive asteroids that are believed to originate in the inner belt, where five distinct sources have been identified: four primitive collisional families (Polana, Erigone, Sulamitis and Clarissa) and a population of low-albedo and low-inclination background asteroids \citep{campins10,campins13,bottke15}. Identifying and characterizing the populations from which these two NEAs might originate will enhance the science return of both missions.

With this main objective in mind we initiated in 2010 a spectroscopic survey in the visible and the near-infrared to characterize the primitive collisional families in the inner belt and the low-albedo background population (PRIMitive Asteroid Spectroscopic Survey, PRIMASS). We started with the largest one, the Polana family \citep{noe15, deleon15}, using, among other, visible spectra obtained with the 10.4m Gran Telescopio Canarias, located at the El Roque de los Muchachos Observatory, in the island of La Palma (Spain). We found that, despite the dynamical and collisional complexity of the Polana family \citep{walsh13,milani14,dykhuis15}, there is a spectral homogeneity both in the visible and the near-infrared wavelengths, all the asteroids showing a continuum in spectral slopes from blue to moderately red typical of B- and C-type primitive asteroids.

To continue with the PRIMASS survey, we have observed and characterized the Erigone family, the second largest one of the four primitive collisional families in the inner belt. We have performed our analysis using data obtained with the 10.4m Gran Telescopio Canarias during the semester 2014B (September 2014 - February 2015). In Section \ref{section2} we describe the observations and the data reduction. In Section \ref{section3} we present the analysis performed on the data, including taxonomical classification, computation of spectral slopes and analysis of aqueous alteration. In Section \ref{section4} we discuss the obtained results, and in Section \ref{section5} we summarize the conclusions.


\section{Observations and data reduction}
\label{section2}

The sample of asteroids we observed in this study has been selected using the Minor Planet Physical Properties Catalogue{\footnote{http://mp3c.oca.eu/MP3C/} (MP$^3$C), which acknowledges as a data source the NASA Planetary Data System (PDS). Orbital data regarding the asteroid families are extracted from a dataset containing asteroid dynamical families including both analytic and synthetic proper elements. These families were computed by David Nesvorny \citep{nesvorny12} using his code based on the Hierarchical Clustering Method (HCM), as described in \cite{zappala90} and \cite{zappala94}. The MP$^3$C catalogue provides also information on the absolute magnitude $H$, the diameter $D$, and the geometric albedo $p_V$. In the case of the diameters and albedos we used the values provided by WISE \citep[Wide-field Infrared Survey Explorer,][]{masiero11}. Family membership is based on values of the synthetic proper elements, i.e., semimajor axis ($a$), eccentricity ($e$) and inclination ($i$), and also on the absolute magnitude as a function of semi major axis ($a,H$). According to these parameters, the Erigone family contains a total of 1785 asteroids.

\begin{figure}
\centering
\includegraphics[width=\hsize]{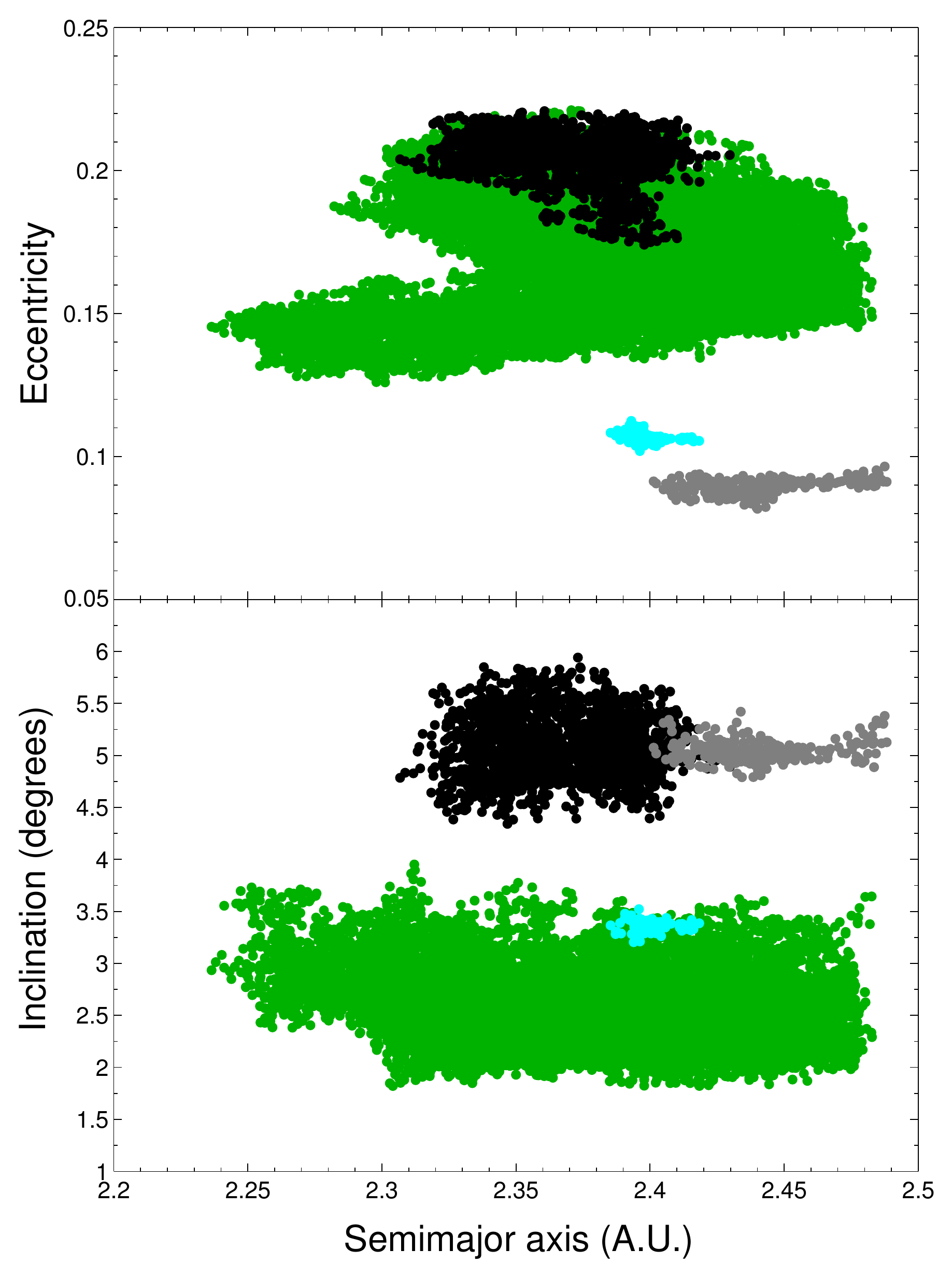}
\caption{Proper semi major axis ($a$) versus proper eccentricity ($e$) and proper inclination ($i$) for the primitive asteroid families in the inner main belt. Erigone family is depicted in black. The Polana (Nysa-Polana complex in this plot), Sulamitis, and Clarissa families are depicted in green, gray, and cyan, respectively.}
\label{fig:proper_elements}
\end{figure}

The selection criterion was quite simple. The Erigone collisional family is a primitive one according to the information we currently have: from the list of 1785 members of the family, 1015 have no albedo information, but from the remaining 770 objects, 692 have geometric albedo values $p_V < 0.1$. Besides, 156 objects have SDSS color-based taxonomies, the majority of them belonging to primitive classes (91 C-types, 10 B-types, 7 X-types, 39 S-types, 5 L-types, 2 K-types, 1 V-type, and 1 A-type). Therefore, we selected those asteroids with geometric albedo $p_V < 0.1$. Observations were done in service mode on different nights (see details in the next section). Therefore, from the previous list of asteroids, we selected those having an apparent visual magnitude in the range $18 < m_V < 21$ to get a good compromise between the number of asteroids observed and the signal-to-noise of the spectra. This means that for some of the nights there were no visible asteroids fulfilling these criteria. In those cases (16 in total), we first tried to find objects having SDSS color-based taxonomical classification available (preferentially C-types). When this information was not available, we selected objects having $p_V > 0.1$. Our last option was to select objects with no information on their albedo. We describe this small sub-sample in detail in Section \ref{section3}.

\subsection{Observations}
We obtained low-intermediate resolution visible spectroscopy for a total of 101 asteroids using the Optical System for Imaging and Low Resolution Integrated Spectroscopy (OSIRIS) camera spectrograph \citep{cepa00,cepa10} at the 10.4m Gran Telescopio Canarias (GTC), located at the El Roque de los Muchachos Observatory (ORM) in La Palma, Canary Islands, Spain. The OSIRIS instrument consists of a mosaic of two Marconi CCD detectors, each with 2048 x 4096 pixels and a total unvignetted field of view of 7.8 x 7.8 arcmin. The single pixel physical size is 15 $\mu$m, giving a plate scale of 0.127 "/pix. To increase the signal to noise for our observations we selected the 2 x 2 binning mode with a readout speed of 200 kHz (which has a gain of 0.95 e-/ADU and a readout noise of 4.5 e-), as corresponds with the standard operation mode of the instrument.

All the spectra were obtained using the OSIRIS R300R grism, which produces a dispersion of 7.74 \AA/pix for a 0.6" slit (in the worst case scenario, a seeing of 3.0" would translate into a resolution of 38.7 \AA/pix), with a spectral coverage from 4800 to 10000 \AA. The R300R grism is used in combination with a second order spectral filter. However, there is still a slight contamination in the spectrum, with a distinguishable contribution for wavelengths at 4800-4900 \AA\ and 9600-9800 \AA. Then, to be conservative, we do not consider here data beyond 9000 \AA. A 5.0"-width slit was used to account for possible variable seeing conditions and it was oriented to the parallactic angle to minimize loses due to atmospheric dispersion. Series of three spectra (whenever possible) were taken for all the targets, with exposure times ranging from 150-600 s, depending on the target brightness. Observational details are listed in Table \ref{table:observations}. Information includes asteroid number, date of observation, starting UT, airmass, exposure time, solar analogue stars and seeing at the moment of the observations. Consecutive spectra were shifted in the slit direction by 10 arcsecs, in order to improve the sky subtraction and the fringing correction.

Observations were done in service mode (within GTC program GTC39-14B) on different nights along September\footnote{Some of the asteroids were observed during August 2014 upon request from the GTC operations team to fill observational gaps during particular nights.}  2014-February 2015. Night conditions were rather variable, covering a wide range of different weather conditions. This was due to the fact that the program was classified as a "filler" (C-band) program within the GTC nightly operation schedule. The aim of these type of programs is to obtain high signal to noise spectra for targets that are relatively bright for a 10m-class telescope as GTC in non-optimal weather conditions, that would include a high seeing value (larger than 1.5 arcsec), bright moon, or some cirrus coverage. Because of this, spectra quality might vary from one night to another (see Table \ref{table:observations}, last column). Since the weather conditions (i.e. clouds, sky brightness, etc.) are the main constraint along the observation, this variation is unrelated to the target brightness. For completeness, we have included in this study the visible spectra of asteroids (163) Erigone, the parent body of the family and (571) Dulcinea, the second largest asteroid in the family, both from the SMASS II catalog\footnote{Available at http://smass.mit.edu/catalog.php}. There were no more asteroids from the Erigone family with published visible spectra. All in all, our sample of asteroids from the Erigone collisional family includes a total of 103 objects.

\subsection{Data reduction}
\label{reduction}

A reduction pipeline for asteroid spectroscopic data obtained with the GTC was developed in order to optimize the reduction process. This pipeline combines standard IRAF\footnote{IRAF is distributed by the National Optical Astronomy Observatories, which are operated by the Association of Universities for Research in Astronomy, Inc., under cooperative agreement with the National Science Foundation.} tasks and some MATLAB functions. 

Using the IRAF tasks included in our pipeline, images were initially bias and flat-field corrected, using lamp flats from the GTC Instrument Calibration Module. Sky background was then subtracted, and a one dimensional spectrum was extracted using an extraction aperture that varied depending on the seeing of the corresponding night. After the extraction, the one dimensional spectra were  wavelength calibrated using Xe+Ne+HgAr lamps. As a final step, the three spectra of the same object, when available, were averaged to obtain one final spectrum of the asteroid.

In order to correct for telluric absorptions and to obtain relative reflectance spectra, at least one solar analogue star from the Landolt catalogue \citep{landolt92} was observed each night. When possible, more than one solar analogue star was observed in order to improve the quality of the final spectra and to minimize potential variations in spectral slope introduced by the use of one single star. These stars were observed using the same spectral configuration as that for the asteroids, and at a similar airmass. The list of the solar analogues used in this study is shown in Table \ref{table:solaranalogues}.

\begin{table}
\caption{Equatorial coordinates of the solar analogue stars used to obtain the reflectance spectra of the observed asteroids.} 
\label{table:solaranalogues} 
\centering 
\begin{tabular}{l l c c} 
\hline\hline 
ID & Star & $\alpha$ & $\delta$ \\ 
\hline 
1 & SA 93-101 & 01:53:18.0 & +00:22:25 \\ 
2 & SA 98-978 & 06:51:34.0 & -00:11:28 \\
3 & SA 102-1081 & 10:57:04.4 & -00:13:10 \\
4 & SA 110-361 & 18:42:45.0 & +00:08:04 \\
5 & SA 112-1333 & 20:43:11.8 & +00:26:15 \\
6 & SA 115-271 & 23:42:41.8 & +00:45:10 \\
7 & SA 107-998 & 15:38:16.4 & +00:15:23 \\
\hline 
\end{tabular}
\end{table}

The MATLAB routines in our pipeline were used to align the spectra of the objects and the corresponding solar analogue, using the theoretical wavelength positions of the telluric lines at 6867.19 and 7593.70 \AA. Once aligned, the spectrum of the object was divided by that of the solar analogue, and the result was normalized to unity at 0.55 $\mu$m. When more than one solar analogue was observed, we divided the spectrum of the asteroid by the spectra of the stars and checked against any possible variations in spectral slopes, which were of the order of 0.6\%/1000 \AA. A variation smaller than 1\%/1000 \AA \ is typically considered as a good value.

Once the final spectrum for each object was obtained, a binning was applied to each spectra, taking intervals of 11 points as the bin size. Then, the reflectance value corresponding to the central wavelength of the bin size was substituted with the median reflectance value, in order to avoid spectral disturbances, and thus making the resulting spectrum more robust. In order to choose the binning size, we selected the worst spectrum in our sample and applied different binning sizes, until its quality improved. Since the finest spectral feature we want to measure, the 0.7 $\mu$m band, has an approximate width of 2000 \AA, we consider that the selected binning is sufficient to improve the quality of the spectra, and to not affect the obtained results. The spectral range extends from 0.5 to 0.9 $\mu$m, with a step of 0.0055 $\mu$m. Spectra are shown in Fig. \ref{fig:final_spectra}.


\section{Analysis and results}
\label{section3}

After computing the final spectra, a taxonomic classification was made using M4AST\footnote{http://m4ast.imcce.fr/}, which is an online tool for modelling of asteroid spectra \citep{popescu12}. The method used by the M4AST tool to classify asteroid spectra is the following: first, the spectrum is fitted with a polynomial curve, and then this curve is compared to each of the classes defined by the \cite{busdemeo2009} taxonomy at the corresponding wavelengths. The tool then selects the taxonomical class producing the smallest standard deviation.

We are dealing with spectra in the visible wavelength range. Thus, we are using the \cite{busbinzel02} taxonomy, in which most of the classes overlap the \cite{busdemeo2009} ones. We have checked individually those cases in which the taxonomical classes were exclusive to the \cite{busdemeo2009} taxonomy and visually classified them according to \cite{busbinzel02}.

To obtain robust results, both original and binned spectra were classified using the M4AST tool. A ${\chi}^2$ method \citep{bevington92} was used to test how well the spectra fitted to the templates. The chosen result was the one corresponding to the smallest standard deviation. Whenever a spectrum was very different from the best-fitting template, or when the standard deviations were very similar for two or more taxonomical classes, we simplified the method: the binned spectrum was fitted to a third-order polynomial and then the procedure was repeated. The taxonomical classification obtained for each asteroid is shown in the last column of Table \ref{table:dynam_info}.


\begin{figure}
\centering
\includegraphics[width=8cm]{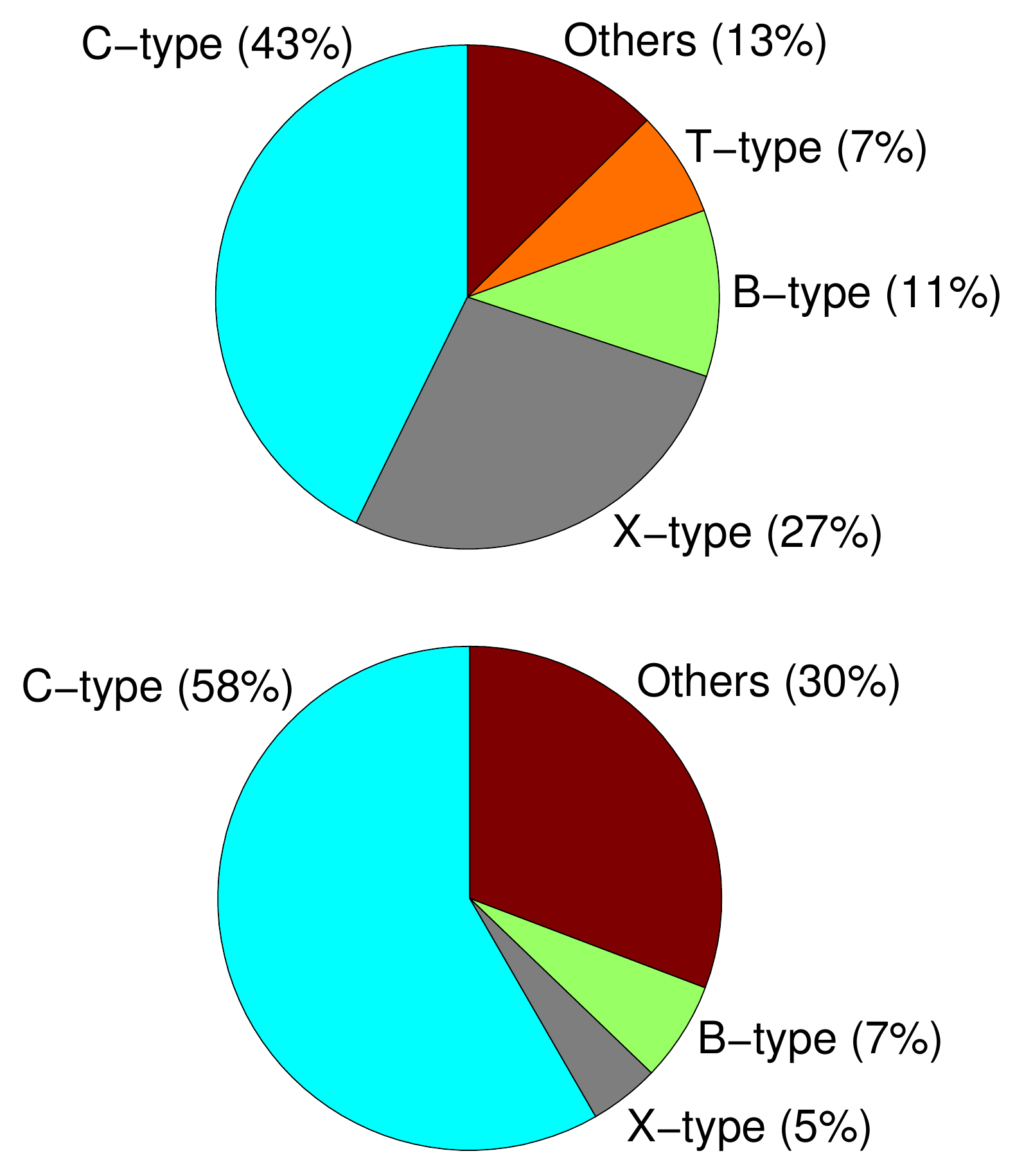}
\caption{Distribution of the taxonomical classes for the Erigone family. Top panel shows the distribution for the sample of 103 asteroids studied in this paper. Bottom panel shows the distribution for the color-based taxonomy from the SDSS data (the X class in the SDSS classification includes the T-type from the Bus taxonomy). The "Others" class includes all non-primitive taxonomies (S-types, V-types, and L-types).}
\label{fig:pie_chart}
\end{figure}

The classification yielded a total of 44 C-type asteroids (including the subclasses Cb, Cg, Ch, and Cgh), 28 X-type asteroids (and subclasses Xc and Xk), 11 B-types, 7 T-types, and 13 objects with non-primitive classifications: 6 S-types (and subclasses), 1 V-type, and 6 L-types. These results are illustrated with a pie chart in Fig. \ref{fig:pie_chart} (upper panel). As expected from our selection criterion (objects with $p_V < 0.1$), the majority of the asteroids belong to primitive taxonomical classes (C-, B-, X-, and T-). As mentioned in Section \ref{section2}, and due to observational constraints, 16 asteroids from the 101 observed did not fulfill the selection criteria. Their taxonomical classification is shown in Table \ref{table:subsample}. Besides, the three asteroids having $p_V > 0.1$ correspond to non-primitive classes. In the case of the remaining 10 asteroids with no albedo information we found a mixture of taxonomies, having 2 C-types, 2 X-types, 2 S-types, 3 L-types, and 1 V-type. From the total of 86 asteroids having $p_V < 0.1$, there is only one single object with a non-primitive classification, being an L-type.

To perform one final comparison with the taxonomical distribution we found from our visible spectra, we searched for all the asteroids in the Erigone family having SDSS color-based taxonomy.  A total of 156 objects belonging to the Erigone family were classified according to \citep{demeo13}. In the lower panel of Fig. \ref{fig:pie_chart} we show their taxonomic distribution. There is a good agreement between the proportion of C and B-types from our sample and the one from the SDSS taxonomy, being the proportion of X-type asteroids significantly larger in our case. The difference between both non-primitive distributions is probably due to the selection criterion ($p_V<0.1$).

\begin{table}
\caption{Asteroids in our sample for which there is no information on their visible geometric albedo or the albedo value is larger than 10\% ($p_V > 0.1$).}
\label{table:subsample}
\centering 
\begin{tabular}{l c c c}  
\hline\hline 
Asteroid & p$_v$ & SDSS Class. & M4AST Class. \\ 
\hline
38661 & - & S & Sr \\ 
39895 & - & S & S \\
56349 & - & C & B \\
186446 & - & C & Ch \\
\hline
18759 & 0.303 & - & Sr\\
24037 & 0.129 & - & L\\
132383 & 0.356 & - & L\\
\hline
38106 & - & - & L \\
50068 & - & - & L \\
66403 & - & - & Ch \\
69706 & - & - & S \\
70511 & - & - & V \\
76922 & - & - & Xk \\
85727 & - & - & S \\
107070 & - & - & L \\
166264 & - & - & Xk \\
186446 & - & - & Ch\\
\hline 
\end{tabular}
\end{table}

Fig. \ref{fig:yorp_cone} shows the distribution in the ($a,H$) space of the 1785 asteroids that have been identified as members of the Erigone collisional family (grey circles). As described by \cite{vokrouhlicky06}, the family shows signs of having experienced dynamical spreading via the Yarkovsky thermal forces. Solid curves in Fig. \ref{fig:yorp_cone} define the boundaries of the family, also known as the ``Yarkovsky cone", computed using the following expression:
\begin{displaymath}
0.2H=\log_{10}(\Delta a/C),
\end{displaymath}
where $\Delta a = a - a_c$, with $a_c$ defined as the center of the family. In practice, $a_c$ is often close to, or the same as, the semimajor axis of the largest member of the family, in this case asteroid (163) Erigone \citep{vokrouhlicky06,bottke15}. \cite{bottke15} showed that, for the Erigone family, $C=1.9$ x $10^{-5}$. This Yarkovsky cone is basically an envelope around the center of the family, indicating the furthest that a family member can drift as a function of its size. Objects outside this cone are likely family interlopers. Fig. \ref{fig:yorp_cone} shows the position, with respect of this cone, of the asteroids studied in this work. Different colors are associated to different spectral classes, as it is indicated in the legend of the figure. It is interesting to note that most of the asteroids having non-primitive taxonomies fall outside the family boundaries, confirming they most likely are interlopers.

In the following sections we will perform a more detailed analysis of the asteroids of the Erigone family with a primitive taxonomical classification, i.e., C-, X-, B-, and T-types. A total of 90 objects have been analyzed.


\begin{figure*}
\centering
\includegraphics[width=0.8\textwidth]{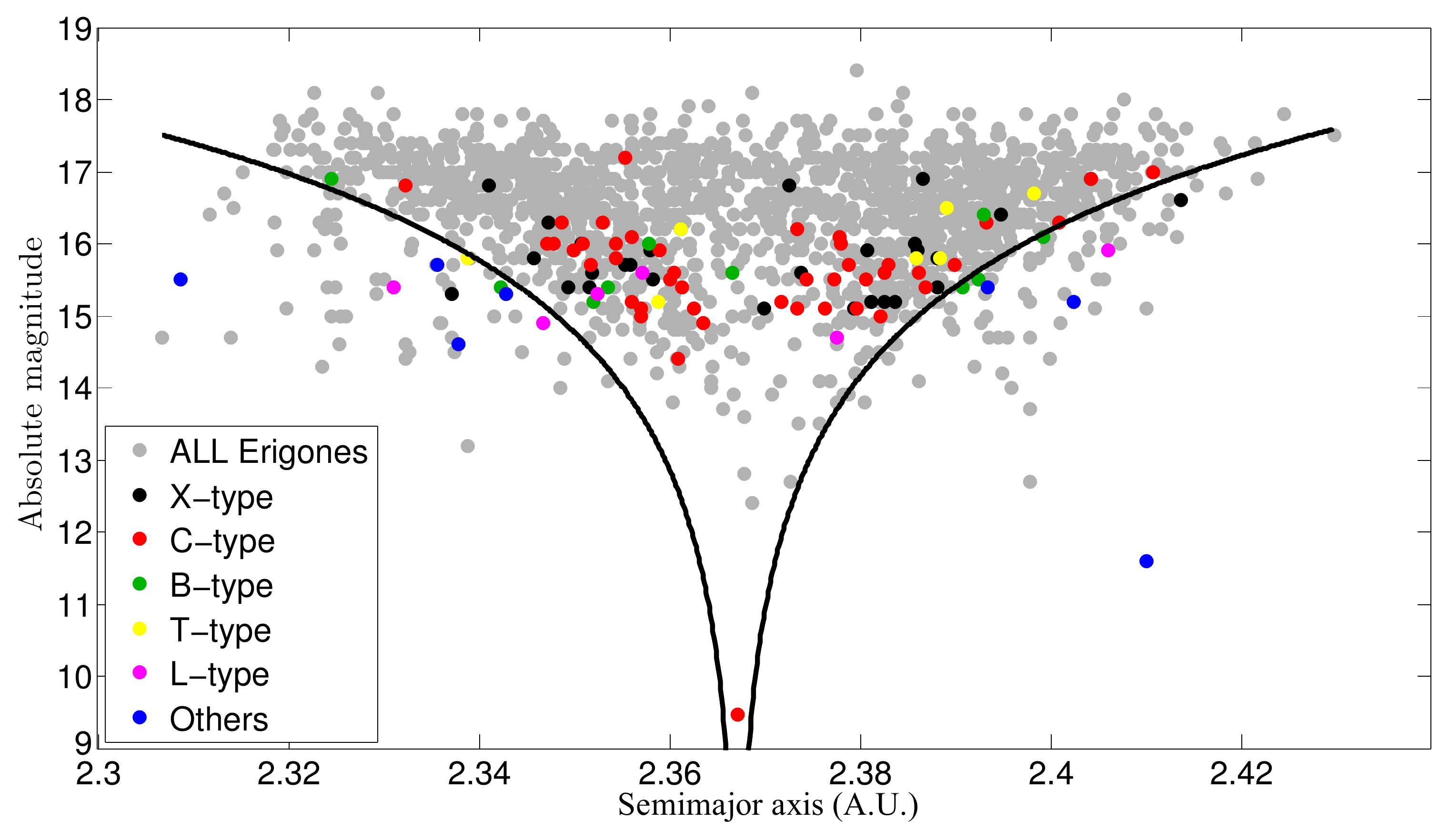}
\caption{Absolute magnitude ($H$) of the asteroids from the Erigone family studied in this work as a function of their proper semimajor axis. The Erigone family (a total of 1785 objects) is shown in grey. The colored circles correspond to the different taxonomical classes found for our sample of 103 members: C-types (red), X-types (black), B-types (green), T-types (yellow), and the rest of non-primitive classes (S-types, L-types, and V-types, in blue). The solid lines represent the boundaries of the family (so called "Yarkovsky cone"). The object having the smallest value of $H$ and located at the bottom of this cone is the parent body of the family, asteroid (163) Erigone.}
\label{fig:yorp_cone}
\end{figure*}

\subsection{Spectral slopes}

Given that primitive asteroids have featureless, linear spectra, we started by computing the spectral slope $S'$, as defined by \citep{luuhewitt90}, between 0.55 and 0.90 $\mu$m:
\begin{displaymath}
S'=\left(\frac{dS/d\lambda}{S_{0.55}}\right),
\end{displaymath}
where $dS/d\lambda$ is the rate of change of the reflectivity in the aforementioned wavelength range, and $S_{0.55}$ is the reflectivity at 0.55 microns. To compute it, a linear least-squares fit between 0.55 and 0.90 $\mu$m has been applied to every primitive asteroid spectra. We normalize\footnote{This is the central wavelength of the Johnson $V$ filter, which is usually used as normalization reference.} the slope at 0.55 $\mu$m. $S'$ is measured in units of $\%/1000$ \AA. The resulting values for the computed spectral slopes are shown in Table \ref{table:results}. The slope errors take into account both the $1\sigma$ uncertainty of the linear fit and the variation of  $0.6\%/1000$ \AA, attributable to the use of different solar analogue stars during the night (see Section \ref{reduction} for more details). Fig. \ref{fig:hist_slopes} shows the distribution of the computed slopes for the 90 primitive objects of the Erigone collisional family (red). As a comparison we show the distribution of the visible spectra slopes of the asteroids of the Polana family (blue) from \cite{deleon15}, which are compatible with a B-type parent. The two distributions are significantly different, with the asteroids of the Erigone family showing, in general, redder spectral slopes. 


\begin{figure}
\includegraphics[width=9cm]{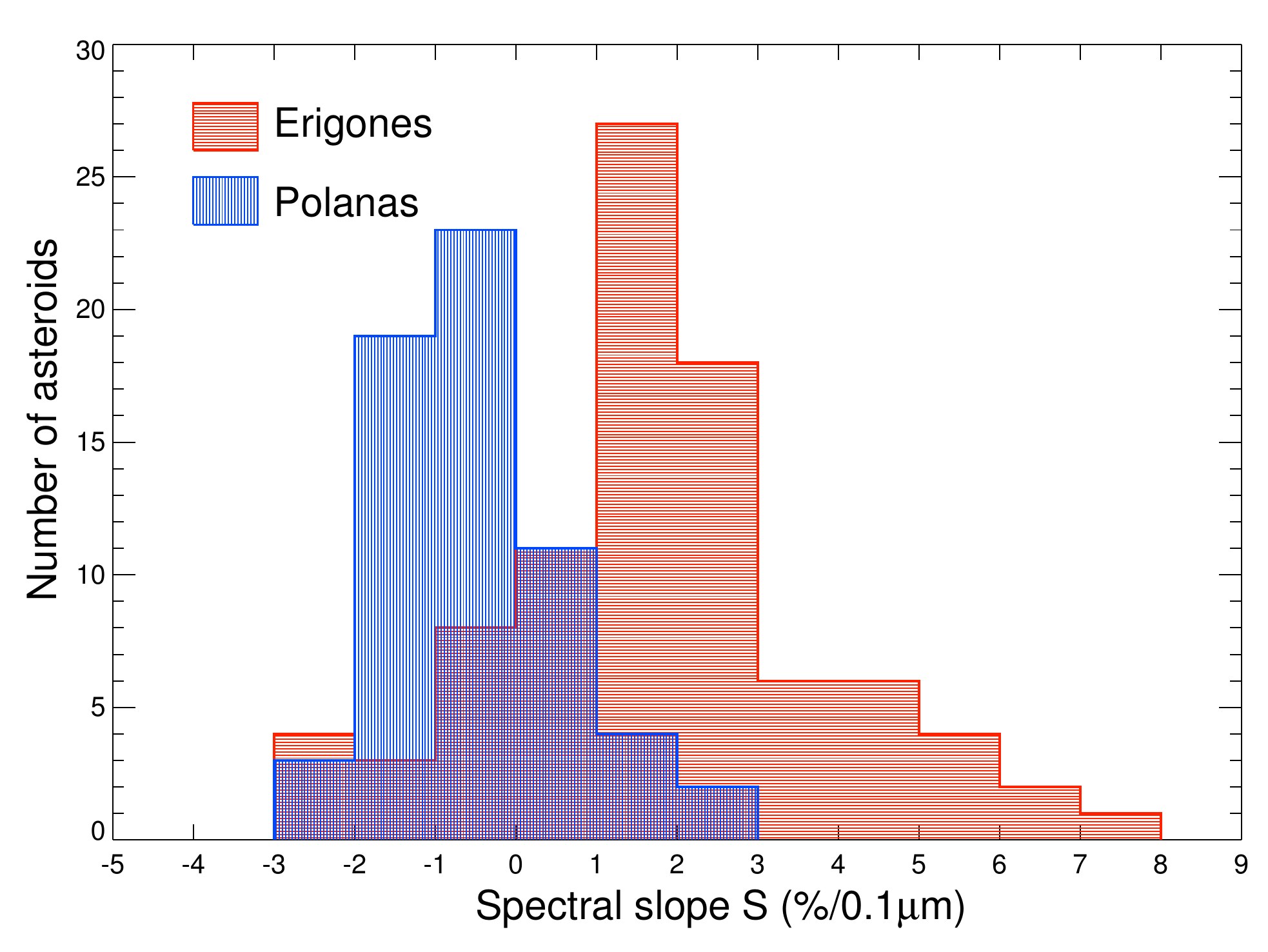}
\caption{Histogram showing the distribution for the values of the spectral slope for the primitive asteroids in the Erigone family (red). The distribution of the slopes of the asteroids of the Polana family from \cite{deleon15} is shown in blue as a comparison.}
\label{fig:hist_slopes}
\end{figure}

\subsection{Aqueous alteration}
\label{subsection:aqueous}

Several studies \citep{vilas94,fornasier99,carvano2003,rivkin12,fornasier14} show that a considerable number of main belt primitive asteroids present an absorption feature around 0.7 $\mu$m, attributed to charge transfer transitions in oxidized iron \citep{vilasgaffey89,vilas94,barucci98}, which is indicative of, or associated with, aqueous alteration in the surface of these objects (i.e. presence of hydrated minerals). \cite{fornasier14} showed that (163) Erigone, the parent body of the family studied in this work,  presents this particular absorption feature at 0.7 $\mu$m, with a depth of $2.2\pm0.1\%$ with respect to the continuum. To search for said absorption feature among our Erigone family members' spectra we follow the procedure described in \cite{carvano2003} with some minor adjustments. For those objects showing this feature we characterize its central wavelength position and depth.

The first step was to compute the continuum of the absorption band by fitting a straight line tangent to the spectra at two positions, 0.54-0.56 $\mu$m and 0.86-0.88 $\mu$m, which are the limits of the 0.7$\mu$m absorption band (green line in the top panel of Fig. \ref{fig:aqueous_example}). We tested slightly different ranges to compute the continuum, finding no significant changes in our results. Then, we fitted the spectrum in this interval using a forth-order spline (red curve in top panel of Fig. \ref{fig:aqueous_example}).

The final step was to remove  the continuum by dividing the spline fit by the straight line we previously obtained (bottom panel of Fig. \ref{fig:aqueous_example}). In order to compute the depth and the central wavelength position of the absorption band and their corresponding errors we run a Monte Carlo model with 1000 iterations, randomly removing 10 points from the spectrum in the range from 0.54 to 0.88 $\mu$m at each iteration, then repeating the above described procedure. The band depth is computed as the difference, in \%, between a reflectance value of 1 and the reflectance value corresponding to the central wavelength position. The final values for the band depth and central wavelength are computed as the mean values obtained for the full Monte Carlo run, and the errors are the corresponding 1$\sigma$ standard deviations. The criterion to decide whether an object showed an aqueous alteration band was to rule out those objects presenting bands with a depth smaller than 1\%, or relative errors in the computation of the depth larger than 15\%. The detection threshold of 1\% correspond to the peak-to-peak scatter in our spectra, which seems to be a better indicator of the spectrum quality than the calculated signal to noise ratio.

We have to note that there are some cases in which the detected bands have higher than expected depths. In those cases (in particular, that of asteroid 210564), large depth values arise from the fact that the absorption band is quite broad, and the limiting regions are narrow and not perfectly defined. In the specific case of asteroid 210564, if we change the continuum-fitting upper limits, the band depth might vary from $\sim9\%$ to $\sim5\%$. This is due to the fact that when moving the upper limit below 0.88 microns, we might fall already inside the band, producing heavy band depth shifts with small error bars. These special cases might reveal the limitations of our data (if we had information over 0.9 $\mu$m, the band limits would be clearer).


\begin{figure}
\resizebox{\hsize}{!}{\includegraphics{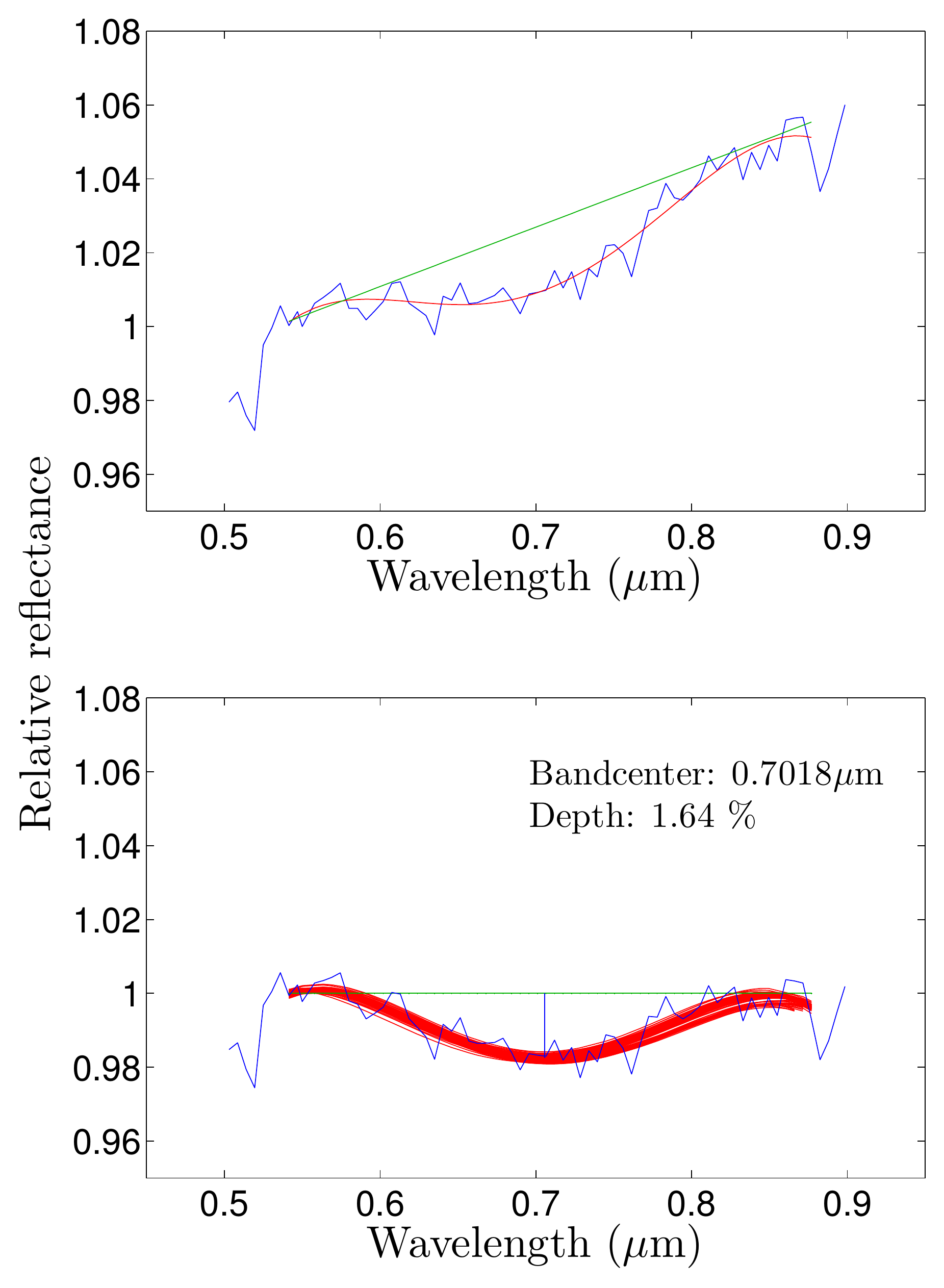}}
\caption{Example figure of the process followed to compute the central wavelength position and the depth of the 0.7 $\mu$m absorption band on the spectra of asteroid (72384). Top panel shows the straight line (green) used to remove the continuum from the fitted absorption band (red curve). Bottom panel shows the result of this continuum removal and the iterative procedure using a Monte Carlo model to compute the mentioned band parameters (see text for details).}
\label{fig:aqueous_example}
\end{figure}

In Table \ref{table:results}, we indicate for each asteroid its taxonomical classification, if it has or not an absorption band at 0.7 $\mu$m (YES/NO), and the center and depth of the band, when present. We found that 52 of the 90 primitive asteroids studied present an absorption band at 0.7 $\mu$m, and that this band is present regardless the specific primitive spectral class. Fig. \ref{fig:hydrated_bars} shows the proportion of asteroids showing hydration band for each primitive class. In the case of C-type asteroids, almost all of them present the hydration feature ($\sim88\%$). This proportion is progressively smaller for B-types ($\sim36\%$), X-types ($\sim28.5\%$), and T-types ($\sim14\%$). This decreasing trend in our results is in agreement with previous studies of the relative incidence of this feature in asteroids distributed through the main belt. \cite{vilas94} found an incidence of 47.7\% in C-type asteroids and 33\% in B-types, while \cite{fornasier14} found an incidence of 50.7\% in C-types and a 9.8\% in B-types. The proportion of C-type asteroids showing the 0.7 $\mu$m absorption band is significantly larger in the case of the Erigone family, with about 87\% of the asteroids.

The presence of a primitive collisional family, with all the observed objects located between 2.3 and 2.4 AU, and with the majority of its members showing the 0.7 $\mu$m hydration band is in good agreement with the results presented by \cite{fornasier14}, where it is suggested that the aqueous alteration processes dominate in primitive asteroids located between 2.3 and 3.1 AU. Moreover, it is stated that the proportion of hydrated primitive objects in the region where the Erigone family is located is 64\%, in very good agreement with the proportion of hydrated objects we have found in our sample, this is, 57.7\% (52 out of 90).

We found no significant correlations between the band depths or the band centers and the taxonomical class, the asteroids orbital parameters, the albedo, or the size of the objects. A similar lack of correlations is found by \cite{carvano2003} and \cite{fornasier14} in their respective studies for asteroids through the whole main belt. Additionally, the mean values we found for both the band depth, $2.9\pm1.5\%$, and the band center position, $7053\pm160\AA$, are in excellent agreement with those obtained by \cite{fornasier14}: $2.8\pm1.2\%$ for the band depth and $6914\pm148\AA$ for the band center. These results suggest that the values obtained for the parameters used to characterize the 0.7 $\mu$m hydration band are independent of the location of the asteroids in the main belt.


\begin{figure}
\includegraphics[width=\hsize]{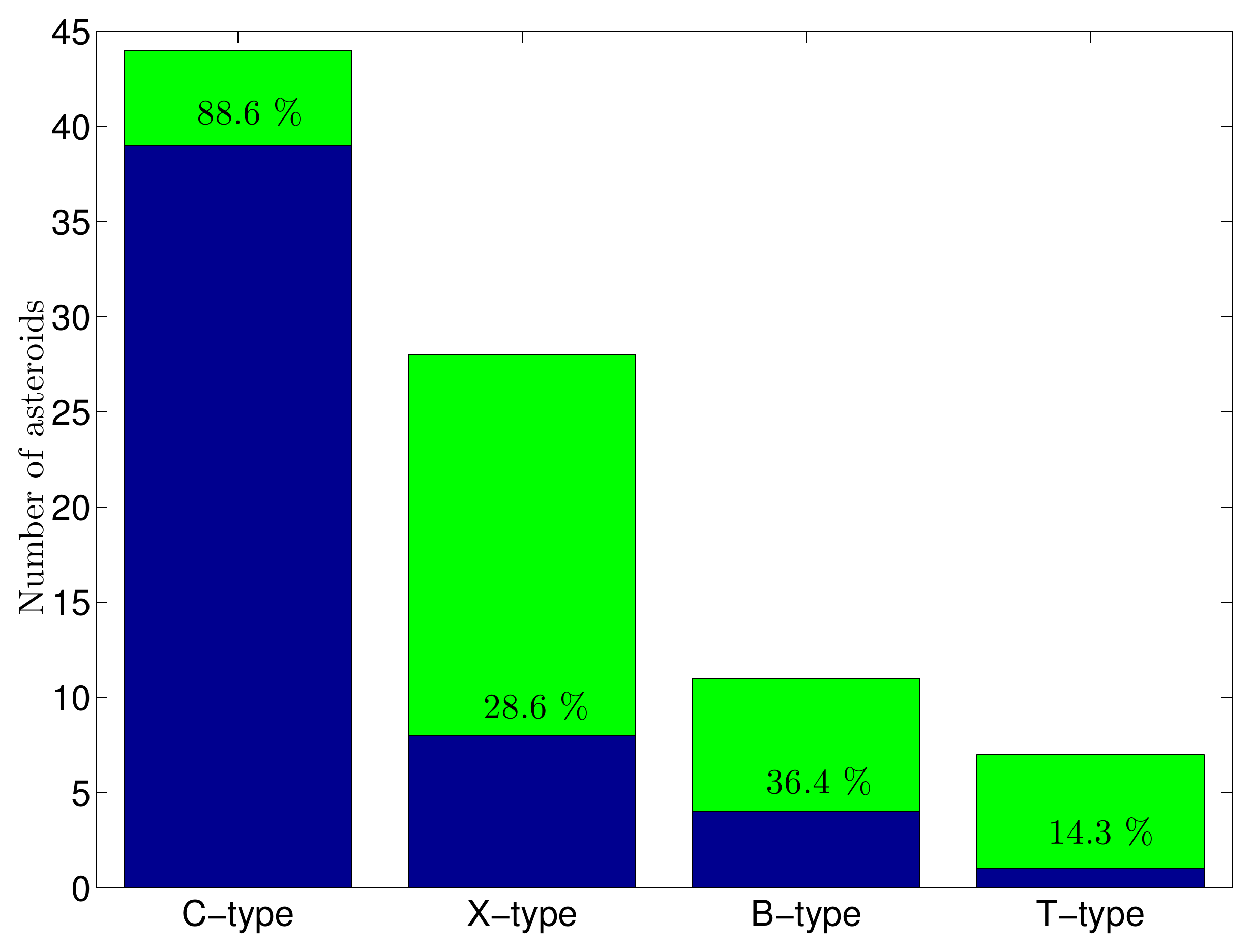}
\caption{Percentage of asteroids showing the 0.7 $\mu$m absorption band (dark blue) for each primitive taxonomic class (green)}.
\label{fig:hydrated_bars}
\end{figure}

\subsection{Comparison with spectra of (101955) Bennu and (162173) Ryugu}

As we described in Section \ref{introduction}, the aim of the characterization of the Erigone primitive family is that, together with the Polana, Clarissa and Sulamitis families, and the low-albedo and low-inclination background asteroids, they are the most likely sources of the two NEAs that are the targets of OSIRIS-REx and Hayabusa 2 sample-return missions: (101955) Bennu and (162173) Ryugu. Spectroscopic and photometric observations of these asteroids suggest that they are composed of primitive materials, pointing to an origin in the aforementioned populations, reinforced by the results of dynamical simulations \citep{campins10,campins13,bottke15}.

Spectral comparison might shed light upon the origins of both asteroids and so, we have compared the available visible spectra of asteroids (101955) Bennu and (162173) Ryugu with the data obtained in this work. 

In the case of (162173) Ryugu, there are several references in the literature showing visible spectra of this small ($\sim$800m), low albedo ($p_V$ = 0.07) near-Earth asteroid. A first spectrum from \cite{binzel01} shows an ultraviolet drop-off in reflectance short-wards 0.65 $\mu$m and provides a classification of Cg-type. Two other visible spectra were presented in \cite{vilas08}, obtained on July and September 2007. These two spectra were different from each other, and also different from the one by \cite{binzel01}, showing no ultraviolet drop-off. The spectrum obtained in July showed an absorption band at 0.7 $\mu$m and a red spectral slope, while the one obtained in September, with a much higher signal-to-noise, presented a neutral slope and showed a marginal, very shallow absorption centered near 0.6 $\mu$m According to \cite{vilas08}, these differences suggest that the surface of the asteroid covers the conjunction of two different geological units. A comparison between these three spectra can be seen in the left panel of Fig. 3 from \cite{campins13}. Additional rotationally resolved visible spectra of (162173) Ryugu were presented by \cite{lazzaro13}, \cite{moskovitz13}, and \cite{sugita13}, all of them compatible with a C-type classification and showing no absorption feature at 0.7$\mu$m and a spectral slope similar to the September 2007 spectrum from \cite{vilas08}, making very unlikely the suggestion of two different surfaces. Therefore, being the one with the highest signal-to-noise, we selected it to perform our spectral comparison. The upper panel of Fig. \ref{fig:ju3_and_bennu} shows the September 2007 \citep{vilas08} spectrum of Ryugu compared to the mean visible spectra of the asteroids of the Erigone family classified as C-types (and subclasses), in red, and those classified as B-types in blue. The visible spectrum of Ryugu is in good agreement with the mean spectrum of C-type asteroids in the Erigone family, even if not showing the 0.7 $\mu$m absorption feature.

Even if the signal-to-noise is quite poor, we have done the exercise of computing both the wavelength central position and the band depth of the 0.7 $\mu$m absorption band present in the July 2007 spectrum of Ryugu \citep{vilas08}. Our calculations yielded a depth of $11.7\pm1.3\%$, and a band center of $6870\pm55$\AA. The absorption band depth and band center are very different from the mean values computed by \cite{fornasier14}, and also from the ones computed in this work for the asteroids in the Erigone family (see section \ref{subsection:aqueous}). This result, together with the absence of the 0.7 $\mu$m absorption band in all the subsequent visible spectra of Ryugu obtained by other authors, suggest that the absorption band observed in the July 2007 spectrum from \cite{vilas08} might be due to some artifact.


\begin{figure}
\resizebox{\hsize}{!}{\includegraphics{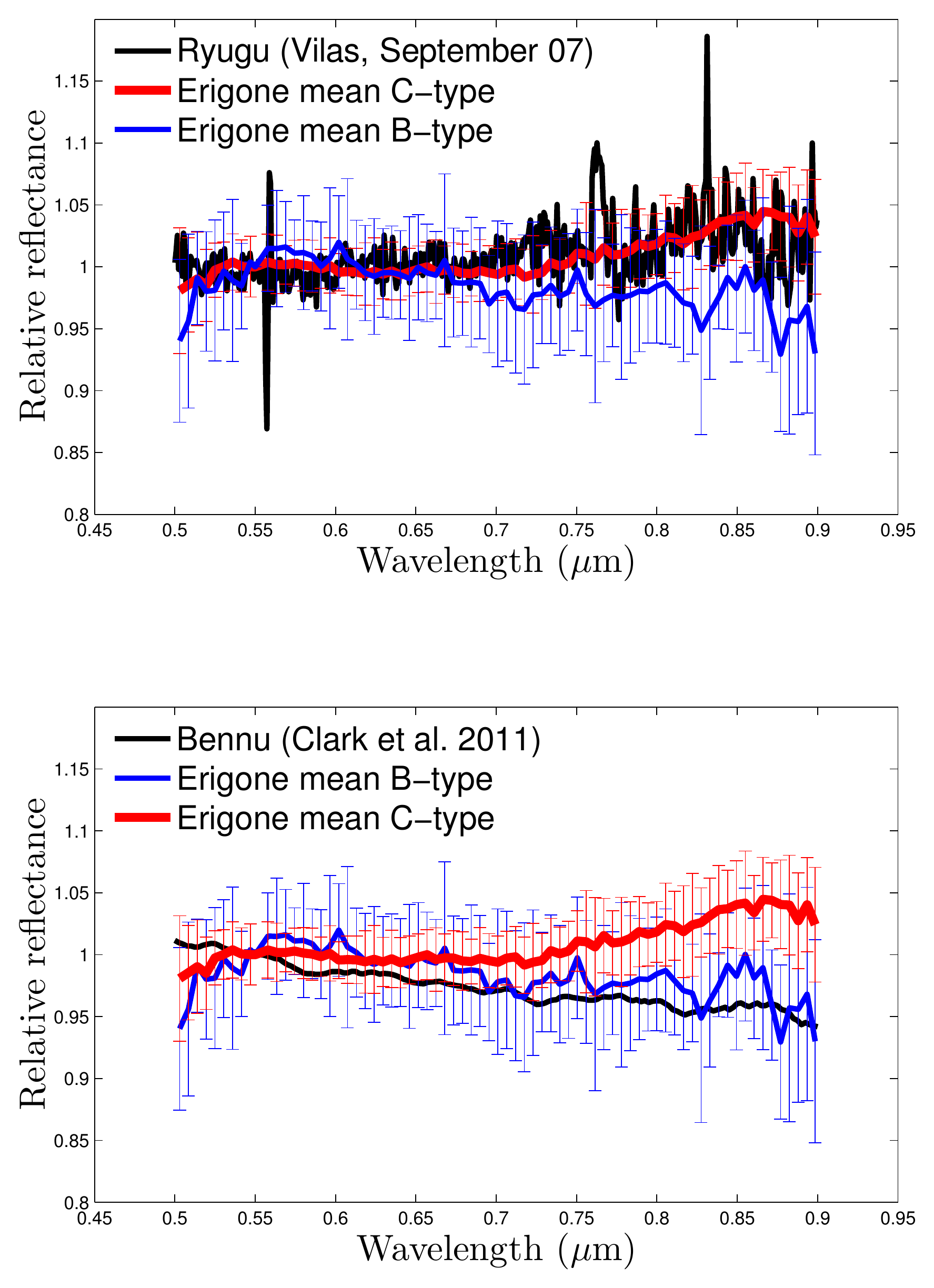}}
\caption{Comparison between the September 2007 \citep{vilas08} visible spectrum of Ryugu in the top panel and the visible spectrum of (101955) Bennu from \cite{clark11} in the bottom panel, with the mean spectra of the C-type (red) and B-type (blue) asteroids in the Erigone family. Standard deviation of the mean ($\pm1\sigma$) is shown with vertical lines.}
\label{fig:ju3_and_bennu}
\end{figure}

Regarding (101955) Bennu, a primitive ($p_V = 0.043$) and rather small ($\sim$500 m) near-Earth asteroid, the only available visible data are found in \cite{clark11} and \cite{hergen13}. According to these studies, Bennu is classified as a B-type asteroid in the \cite{busbinzel02} taxonomy. In addition, in \cite{binzel15}, spectra obtained in the NIR range show spectral variability, pointing towards a C-type class according to \cite{busdemeo2009}. As it has been shown in \cite{clark10} and \cite{deleon12}, asteroids classified as B-types in the visible can present considerable slope variation in the NIR, from negative, blue slopes, to positive, redder ones. The lower panel of Fig. \ref{fig:ju3_and_bennu} shows the visible spectrum of Bennu (black) compared to the mean visible spectra of the asteroids of the Erigone family classified as C-types (and subclasses), in red, and those classified as B-types in blue, observed in this work. From a visual inspection, the visible spectrum of Bennu seems to marginally show the presence of the hydration feature 0.7 $\mu$m. We followed the same approach as in the previous section in order to study the presence of a possible aqueous alteration band. Our calculations yielded the presence of a band centered at 7486$\pm$76 \AA, with a depth of 0.96$\pm$0.02$\%$. This depth is slightly below the threshold (1\%) established for a positive detection, and the wavelength position of the band center is significantly different from the mean value found for the family ($7053\pm160\AA$). Therefore, we rule out the presence of this hydration band in the spectrum of Bennu.


\section{Discussion}
\label{section4}

No previous spectroscopic studies had been performed until now on the Erigone primitive collisional family, with only two asteroids already classified using visible spectroscopy: (163) Erigone, and (571) Dulcinea. It is clearly shown that the asteroids studied in this work are spectroscopically consistent with the hypothesis of a common parent body, and that this parent body is (163) Erigone, classified as a C-type asteroid. Moreover, (163) Erigone shares the particular spectroscopic feature at 0.7 $\mu$m with most of the family members. Another primitive collisional family in the inner belt, the Polana family, that, together with Erigone, Sulamitis, and Clarissa, are the four primitive families in the inner belt, has been recently studied by several authors \citep{walsh13,milani14,dykhuis15,deleon15,noe15}. Comparing our results on the Erigone family with those obtained from \cite{deleon15} for the Polana family, we observe two main differences:
\begin{itemize}
\item[a)] The Erigone family presents a different distribution of taxonomical classes from that of the Polana family, referred to as the Polana-Eulalia complex in \cite{deleon15}. In our sample, we found 44 C-type objects, 28 X-types, 11 B-types and 7 T-types, plus 13 interlopers (S-types and L-types). The mean spectra for each class are clearly differentiated, specially in the case of the X and T classes (see Fig. \ref{fig:mean_classes}). On the contrary, in the Polana family we found mainly C-types (51\%) and B-types (42\%), with few X-types (5\%) and only one S-type, with the mean spectra for the C-, B-, and X-types presenting similar values \citep{deleon15}. In addition, the slope distribution for the objects in the Erigone family is fairly redder than that of the Polana family, as can be seen in Fig. \ref{fig:hist_slopes}, mainly due to the larger fraction of X-types found in the former. 
\newline
\item[b)] The majority of the primitive asteroids in the Erigone family (52 out of 90) show evidence of aqueous altered minerals on their surfaces. We conducted the same analysis as that described in Section \ref{subsection:aqueous} for the data in \cite{deleon15}, and we found that, according to our criterion, only one object in the Polana family (asteroid 29626), showed the 0.7 $\mu$m absorption feature, having a band depth of $1.15\pm0.11\%$ and a band center of $7300\pm211 \AA$. This difference in the two families is well explained by their two parent bodies. Asteroid (142) Polana shows no signs at all of the hydration feature at 0.7 $\mu$m. On the other hand, (163) Erigone, the parent body of the Erigone family, shows evidence of aqueous alteration. Therefore, the presence of the hydration feature in the spectra of most of its family members is somewhat expected. An interesting explanation for the difference in hydration between the two families is presented by \cite{matsukoa15}. They propose that space weathering effects on C-type asteroids tend to shallow the 0.7 $\mu$m absorption feature. Space weathering processes might have removed the aqueous alteration band in the Polana family, since the Polana family is older than Erigone: according to \cite{bottke15}, Erigone is $130\pm30$ Myr old, while Polana (referred to as New Polana in their paper) and Eulalia are $1400\pm150$ Myr old, and $830^{+370}_{-100}$ Myr old respectively.


\begin{figure}
\resizebox{\hsize}{!}{\includegraphics{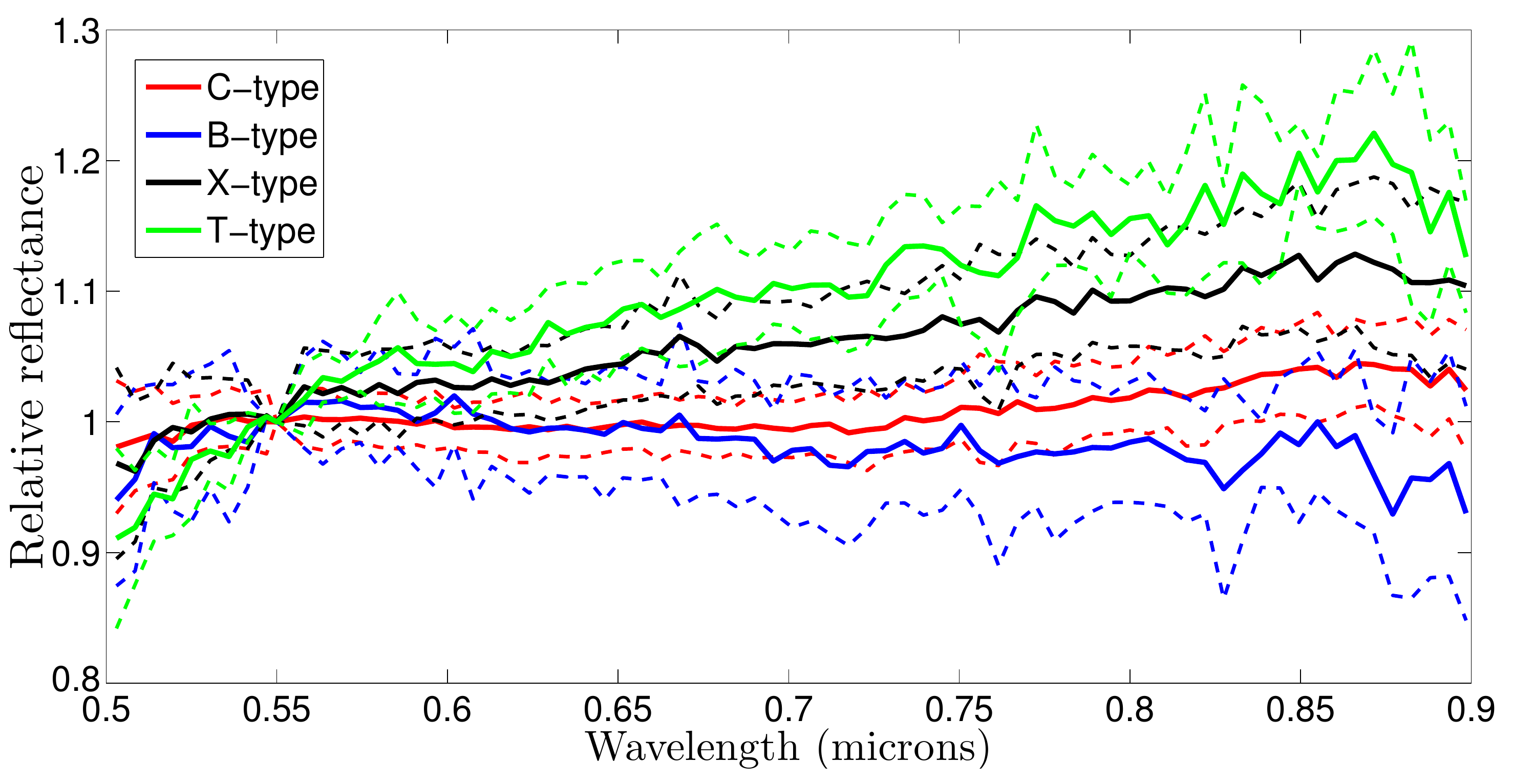}}
\caption{Computed mean spectra of the asteroids in the Erigone family classified as C-types (red), B-types (blue), X-types (black), and T-types (green). The mean spectrum for each class is plotted as a thick line, while the $\pm$1$\sigma$ of the mean is shown with dashed lines. Note that the different mean spectra are well differentiated from one another.}
\label{fig:mean_classes}
\end{figure}

\end{itemize}

Regarding the likelihood of the Erigone family of being the source for the asteroids (101955) Bennu and (162173) Ryugu, we have compared visible spectra of both objects with the mean spectra of the primitive asteroids in the family. In our sample, we have found approximately 56\% of non-interlopers showing evidence of aqueous alteration. According to \cite{bottke15}, there is little to no chance for the smaller families (Erigone, Clarissa and Sulamitis) to be the source of both asteroids:
\begin{itemize}
\item[-] \textbf{(101955) Bennu}: Bennu is classified as a B-type asteroid in the visible wavelength range, according to \cite{clark11} and \cite{hergen13}. Even if the fraction of B-type asteroids in the Erigone family is small, and the probability of Bennu coming from the Erigone family is small \citep{bottke15}, the spectral comparison between the spectrum of Bennu and the mean spectrum of the B-type asteroids is compatible with a possible origin in the Erigone family.
\newline
\item[-] \textbf{(162173) Ryugu}: In their paper, \cite{vilas08} report the presence of an aqueous alteration band around 0.7 $\mu$m in one of the three visible spectra obtained for Ryugu (July 2007, see their Fig. 4). Following the same approach we used for our aqueous alteration study, the depth of the band we computed in the July 2007 spectrum from \cite{vilas08}, $11.7\pm1.3\%$, is far from the medium depth of the aqueous altered asteroids in \cite{fornasier14} ($2.8\pm1.2\%$) and also in this work ($2.9\pm1.5\%$). However, from the spectral comparison, the possibility of JU$_3$ originating in the Erigone family should not be discarded. Ryugu is a C-type asteroid, and the most abundant spectral classes in the Erigone family are C-type asteroids and subclasses (Cb,Cg,Ch and Cgh). This would point to a possible origin of Ryugu in the Erigone family, even if all the visible spectra of this asteroid obtained by different authors show no 0.7 $\mu$m absorption band.
\end{itemize}

Since the Polana family, suggested in \cite{bottke15} as the most probable source for both NEAs, does not show signs of aqueous altered asteroids \citep{walsh13,dykhuis15,deleon15}, the Erigone family cannot be discarded as a possible source for both NEAs. Further observations of (101955) Bennu and (162173) Ryugu should be done in order to check whether the 3 $\mu$m absorption feature is present or not, definitely indicating evidence of aqueous alteration, and confirming or ruling out the chances for Erigone being the source of these objects.

\section{Conclusions}
\label{section5}

We studied a total of 103 visible spectra of asteroids in the Erigone primitive family. We observed 101 of these asteroids with the 10.4m Gran Telescopio Canarias. These objects had never been observed before via spectroscopy. We added the visible spectra of (163) Erigone and (571) Dulcinea from the SMASS database for completeness.

We have seen that the slope distribution and the taxonomic classification of the asteroids studied are in agreement with the assumption of a primitive family. Most of the observed objects are C-types, highly consistent with the classification of (163) Erigone, supporting its status as the parent body of the family. In terms of taxonomical distribution, 42\% of the asteroids in our survey are classified as C-type objects, 28\% as X-types, 11\% as B-types, 7\% as T-types, and 13\% of the objects as non-primitive types, most likely interlopers (S-types and subclasses, and L-types). 

In addition, the study on aqueous alteration performed on the group of primitive objects shows a high number of hydrated asteroids. The C-type class is the one which shows the largest number of hydrated asteroids, namely $\sim$86\%. The other primitive classes, also showing hydration band, present a much smaller fraction: B-type objects, $\sim$36\%, X-type objects, $\sim$28\% and T-type, $\sim14$\%. This distribution of the hydration as a function of the taxonomical class is in agreement with the one in \cite{fornasier14} for asteroids in the main belt. 

Based on the spectral comparison only, and as in the case of the Polana family, studied in \cite{deleon15}, we cannot discard the possibility of Erigone being the source family for near Earth asteroids (101955) Bennu and (162173) Ryugu. The spectral classes present in the family are compatible with the taxonomic classification of both asteroids. Future research should include further spectroscopic study, both in the visible and near-infrared regions, of the other primitive families in the inner belt, such as Sulamitis and Clarissa, in order to completely rule them out as the possible sources of asteroids (101955) Bennu and (162173) Ryugu. 


\begin{acknowledgements}
{\it DM} gratefully acknowledges the Spanish Ministry of Economy and Competitiveness (MINECO) for the financial support received in the form of a Severo-Ochoa PhD fellowship, within the Severo-Ochoa International PhD Program. {\it DM, JdL, JL}, and {\it VL} acknowledge support from the project AYA2012-39115-C03-03 and ESP2013-47816-C4-2-P (MINECO). {\it JdL} acknowledges support from the Insituto de Astrof\'isica de Canarias. {\it HC} acknowledges support from NASA's Near-Earth Object Observations program and from the Center for Lunar and Asteroid Surface Science funded by NASA's SSERVI program at the University of Central Florida. The authors gratefully acknowledge the reviwer, Sonia Fornasier, for her comments and suggestions. The results obtained in this paper are based on observations made with the Gran Telescopio Canarias (GTC), installed in the Spanish Observatorio del Roque de los Muchachos of the Instituto de Astrof\'isica de Canarias, in the island of La Palma.
\end{acknowledgements}



\begin{longtab}
\begin{longtable}{ccccccc}
\caption{\label{table:observations} Observational circumstances of the asteroids presented in this paper. Check Table \ref{table:solaranalogues} for the ID number of each solar analogue star.}\\
\hline\hline
Object & Date & UT start & Airmass & Exposure time (s) & SAs & Seeing (")\\
\hline
\endfirsthead
\caption{Continuation}\\
\hline\hline
Object & Date & UT start & Airmass & Exposure time (s) & SAs & Seeing (")\\
\hline
\endhead
\hline\\
\endfoot
  10992 & 2014-08-16 & 05:46 & 1.117 & 3x200 & 5 & 0.9\\
  11856 & 2015-01-12 & 01:10 & 1.482 & 3x250 & 2 & 2.0\\
  18759 & 2014-12-17 & 22:35 & 1.149 & 3x250 & 1 & 2.0\\
  19415 & 2014-09-18 & 22:23 & 1.310 & 3x300 & 4,6 & 1.0\\
  20992 & 2014-12-17 & 03:13 & 1.248 & 3x250 & 2 & 1.9\\
  23397 & 2015-01-14 & 03:51 & 1.292 & 3x300 & 1,3 & 1.5\\
  24037 & 2014-12-18 & 01:27 & 1.305 & 2x200 & 1 & 2.7\\
  25381 & 2014-12-18 & 00:22 & 1.352 & 3x500 & 1 & 2.0\\
  37437 & 2014-12-17 & 22:04 & 1.335 & 3x500 & 1 & 2.5\\
  38106 & 2014-10-14 & 03:43 & 1.033 & 3x400 & 1,2 & 1.0\\
  38173 & 2014-12-15 & 23:46 & 1.554 & 3x300 & 1,2 & 2.0\\
  38661 & 2014-09-13 & 03:04 & 1.257 & 3x200 & 1,6 & 0.9\\
  39694 & 2014-09-19 & 00:18 & 1.198 & 3x400 & 4,6 & 0.9\\
  39895 & 2014-09-15 & 03:38 & 1.507 & 3x300 & 1,6 & 0.8\\
  42155 & 2014-09-13 & 03:36 & 1.095 & 3x200 & 1,6 & 0.8\\
  42552 & 2014-09-15 & 02:50 & 1.513 & 3x250 & 1,6 & 0.9\\
  44766 & 2014-12-17 & 03:58 & 1.039 & 3x200 & 2 & 1.3\\
  44942 & 2015-01-14 & 03:29 & 1.496 & 3x200 & 1,3 & 1.5\\
  45357 & 2015-01-14 & 04:22 & 1.416 & 3x300 & 1,3 & 1.5\\
  49731 & 2015-01-14 & 04:46 & 1.920 & 3x200 & 1,3 & 1.9\\
  49859 & 2015-05-04 & 23:37 & 1.300 & 3x180 & 3,7 & 1.5\\
  50068 & 2015-05-05 & 00:10 & 1.131 & 3x180 & 3,7 & 1.5\\
  52870 & 2015-01-19 & 05:57 & 1.124 & 3x500 & 3 & 2.0\\
  52891 & 2015-01-15 & 03:53 & 1.197 & 3x500 & 3 & 1.2\\
  56349 & 2015-01-15 & 04:29 & 1.163 & 3x500 & 3 & 1.1\\
  65354 & 2014-12-13 & 01:05 & 1.048 & 3x500 & 2 & 1.5\\
  66309 & 2014-10-14 & 02:34 & 1.456 & 3x250 & 1,2 & 1.4\\
  66325 & 2014-10-13 & 04:20 & 1.847 & 3x300 & 1,2 & 1.5\\
  66403 & 2014-09-14 & 01:35 & 1.110 & 3x200 & 1,6 & 0.9\\
  67891 & 2014-09-15 & 02:09 & 1.684 & 3x200 & 1,6 & 1.0\\
  67918 & 2014-09-13 & 02:39 & 1.214 & 3x200 & 1,6 & 0.9\\
  67940 & 2014-09-15 & 03:14 & 1.709 & 3x200 & 1,6 & 0.9\\
  68114 & 2014-09-17 & 01:22 & 1.142 & 3x250 & 6 & 1.0\\
  68685 & 2014-10-12 & 00:30 & 1.406 & 4x250 & 6 & 0.9\\
  69266 & 2014-12-18 & 01:52 & 1.218 & 3x250 & 1 & 3.0\\
  69706 & 2014-09-15 & 04:56 & 1.023 & 3x500 & 1,6 & 0.8\\
  70312 & 2014-09-18 & 03:01 & 1.088 & 3x300 & 1,4 & 1.3\\
  70361 & 2014-12-17 & 03:38 & 1.180 & 3x200 & 2 & 1.7\\
  70427 & 2015-01-14 & 05:06 & 1.606 & 3x250 & 1,3 & 1.7\\
  70511 & 2014-09-15 & 04:20 & 1.010 & 3x300 & 1,6 & 0.7\\
  71932 & 2014-10-13 & 02:22 & 1.176 & 3x200 & 1,2 & 1.2\\
  72047 & 2014-11-25 & 22:13 & 1.027 & 3x350 & 1 & 1.3\\
  72143 & 2014-09-19 & 23:39 & 1.414 & 3x250 & 1,6 & 1.3\\
  72230 & 2014-09-19 & 01:40 & 1.308 & 2x300 & 4,6 & 0.9\\
  72292 & 2014-12-18 & 00:57 & 1.276 & 3x450 & 1 & 2.0\\
  72308 & 2014-08-16 & 04:04 & 1.304 & 3x250 & 5 & 0.7\\
  72384 & 2014-11-25 & 21:49 & 1.201 & 3x250 & 1 & 1.4\\
  72941 & 2014-10-13 & 03:57 & 1.997 & 3x300 & 1,2 & 1.6\\
  73860 & 2014-09-14 & 03:27 & 1.177 & 3x250 & 1,6 & 0.9\\
  74755 & 2014-11-25 & 22:53 & 1.017 & 3x300 & 1 & 1.2\\
  74962 & 2014-12-02 & 04:21 & 1.136 & 3x500 & 2 & 1.4\\
  75089 & 2015-01-19 & 04:48 & 1.096 & 3x350 & 3 & 2.5\\
  76922 & 2014-10-08 & 00:02 & 1.292 & 6x250 & 1,6 & 0.7\\
  77421 & 2014-10-10 & 05:44 & 1.040 & 2x500 & 1,2 & 0.8\\
  78069 & 2014-12-02 & 05:29 & 1.377 & 3x500 & 2 & 1.7\\
  78826 & 2014-09-13 & 04:53 & 1.114 & 3x200 & 1,6 & 0.8\\
  78889 & 2014-10-12 & 02:05 & 2.199 & 4x150 & 6 & 1.4\\
  79044 & 2014-10-10 & 05:05 & 1.078 & 3x400 & 1,2 & 0.7\\
  85727 & 2015-05-04 & 23:47 & 1.176 & 3x250 & 3,7 & 0.9\\
  96405 & 2014-09-17 & 22:15 & 1.294 & 5x400 & 1,4 & 1.3\\
  96463 & 2014-10-13 & 02:48 & 1.862 & 3x400 & 1,2 & 1.2\\
  96768 & 2014-10-07 & 23:23 & 1.315 & 3x300 & 1,6 & 0.8\\
  98345 & 2014-09-17 & 21:39 & 1.392 & 3x400 & 1,4 & 1.3\\
  100784 & 2014-09-18 & 04:59 & 1.049 & 5x500 & 1,4 & 2.0\\
  106794 & 2014-12-16 & 00:11 & 1.602 & 3x300 & 1,2 & 3.0\\
  107070 & 2014-10-13 & 03:29 & 1.873 & 3x400 & 1,2 & 1.3\\
  107742 & 2014-12-17 & 23:13 & 1.214 & 3x500 & 1 & 2.4\\
  111789 & 2014-09-20 & 01:19 & 1.283 & 3x400 & 1,6 & 1.1\\
  121096 & 2014-10-07 & 22:20 & 1.458 & 3x300 & 1,6 & 0.7\\
  129818 & 2014-10-10 & 04:31 & 1.098 & 3x500 & 1,2 & 0.7\\
  132056 & 2014-10-12 & 02:33 & 1.306 & 3x250 & 6 & 1.1\\
  132383 & 2014-10-14 & 04:39 & 1.026 & 3x400 & 1,2 & 1.0\\
  133123 & 2014-10-07 & 22:51 & 1.329 & 3x400 & 1,6 & 0.7\\
  133197 & 2014-10-12 & 01:37 & 1.954 & 3x250 & 6 & 1.0\\
  133503 & 2014-10-14 & 03:06 & 1.047 & 3x500 & 1,2 & 1.4\\
  137397 & 2014-10-14 & 04:10 & 1.057 & 3x300 & 1,2 & 1.1\\
  162795 & 2014-10-12 & 01:06 & 1.613 & 3x300 & 6 & 1.0\\
  165536 & 2014-12-18 & 02:17 & 1.235 & 3x250 & 1 & 0.7\\
  166264 & 2014-09-18 & 23:28 & 1.178 & 3x400 & 4,6 & 1.0\\
  169066 & 2014-12-16 & 00:42 & 1.301 & 3x300 & 1,2 & 1.8\\
  170184 & 2014-11-25 & 23:23 & 1.190 & 3x400 & 1 & 1.7\\
  174594 & 2014-12-15 & 22:55 & 1.049 & 3x250 & 1,2 & 1.6\\
  175811 & 2014-12-14 & 02:57 & 1.181 & 3x300 & 2 & 2.0\\
  177258 & 2014-12-07 & 06:31 & 1.197 & 3x300 & 2 & 0.9\\
  178844 & 2014-12-16 & 04:41 & 1.210 & 3x350 & 1,2 & 1.6\\
  186446 & 2015-01-21 & 07:01 & 1.163 & 2x600 & 3 & 1.8\\
  186714 & 2014-12-14 & 05:32 & 1.574 & 2x350 & 2 & 2.3\\
  208048 & 2015-01-20 & 00:28 & 1.224 & 3x600 & 1,2 & 2.0\\
  210564 & 2015-01-19 & 23:35 & 1.515 & 3x500 & 1,2 & 2.5\\
  213825 & 2014-12-02 & 03:35 & 1.158 & 3x300 & 2 & 1.4\\
  232922 & 2014-09-18 & 23:01 & 1.177 & 3x200 & 4,6 & 1.0\\
  242324 & 2015-01-16 & 23:07 & 1.212 & 3x500 & 2 & 2.0\\
  243648 & 2014-12-16 & 04:06 & 1.193 & 3x500 & 1,2 & 1.6\\
  250431 & 2015-01-19 & 22:48 & 1.259 & 3x600 & 1,2 & 3.0\\
  251796 & 2015-01-26 & 04:20 & 1.041 & 3x600 & 3 & 1.5\\
  252953 & 2015-01-26 & 03:16 & 1.012 & 3x450 & 3 & 1.5\\
  253538 & 2015-01-19 & 22:02 & 1.304 & 3x500 & 1,2 & 2.5\\
  253798 & 2015-01-26 & 03:47 & 1.031 & 3x400 & 3 & 1.8\\
  256789 & 2014-12-14 & 03:27 & 1.127 & 3x400 & 2 & 2.0\\
  262102 & 2015-01-26 & 05:37 & 1.111 & 3x600 & 3 & 1.6\\
  265259 & 2015-01-26 & 04:59 & 1.109 & 3x600 & 3 & 1.5\\
\end{longtable}
\end{longtab}


\begin{longtab}
\begin{longtable}{cccccccccc}
\caption{\label{table:dynam_info} Physical parameters of the targets}\\
\hline\hline
Object & $a$ (AU) & $e$ & $i$ ($^{\circ}$) & $H_V$ & D (km) & err$_D$ & p$_V$ & err$_{pV}$ & Class\\
\hline
\endfirsthead
\caption{Continuation}\\
\hline\hline
Object & $a$ (AU) & $e$ & $i$ ($^{\circ}$) & $H_V$ & D (km) & err$_D$ & p$_V$ & err$_{pV}$ & Class\\
\hline
\endhead
\hline\\
\endfoot
  163 & 2.367 & 0.21 & 4.801 & 9.5 & 81.58 & 3.06 & 0.033 & 0.004 & Ch\\
  571 & 2.41 & 0.213 & 5.222 & 11.6 & 11.70 & 0.09 & 0.298 & 0.049 & S\\
  10992 & 2.363 & 0.205 & 5.022 & 15.1 & 5.48 & 0.28 & 0.062 & 0.014 & Cgh\\
  11856 & 2.384 & 0.201 & 4.735 & 15.2 & 4.20 & 0.44 & 0.093 & 0.029 & X\\
  18759 & 2.338 & 0.205 & 4.641 & 14.6 & 2.90 & 0.13 & 0.303 & 0.048 & Sr\\
  19415 & 2.382 & 0.209 & 5.055 & 15.0 & 5.76 & 0.18 & 0.058 & 0.005 & C\\
  20992 & 2.337 & 0.199 & 5.156 & 15.3 & 4.02 & 0.38 & 0.074 & 0.035 & Xk\\
  23397 & 2.357 & 0.218 & 5.692 & 15.1 & 7.75 & 3.34 & 0.027 & 0.021 & Ch\\
  24037 & 2.378 & 0.198 & 5.568 & 14.7 & 3.87 & 0.58 & 0.129 & 0.031 & L\\
  25381 & 2.395 & 0.202 & 4.751 & 16.4 & 4.33 & 0.17 & 0.028 & 0.005 & X\\
  37437 & 2.349 & 0.215 & 5.06 & 15.4 & 5.30 & 0.78 & 0.043 & 0.026 & Xc\\
  38106 & 2.331 & 0.213 & 4.781 & 15.4 & - & - & - & - & L\\
  38173 & 2.355 & 0.212 & 4.736 & 15.7 & 3.56 & 0.24 & 0.080 & 0.016 & X\\
  38661 & 2.309 & 0.203 & 4.834 & 15.5 & - & - & - & - & Sr\\
  39694 & 2.358 & 0.208 & 5.369 & 16.0 & 3.97 & 1.20 & 0.041 & 0.035 & B\\
  39895 & 2.393 & 0.198 & 5.526 & 15.4 & - & - & - & - & S\\
  42155 & 2.38 & 0.203 & 5.225 & 15.1 & 5.93 & 0.93 & 0.049 & 0.018 & Cgh\\
  42552 & 2.356 & 0.208 & 4.938 & 15.2 & 4.90 & 0.33 & 0.051 & 0.014 & Cgh\\
  44766 & 2.363 & 0.21 & 5.207 & 14.9 & 6.46 & 0.76 & 0.041 & 0.022 & Cgh\\
  44942 & 2.383 & 0.204 & 5.215 & 15.2 & 4.83 & 0.67 & 0.063 & 0.051 & X\\
  45357 & 2.374 & 0.215 & 5.211 & 15.5 & 3.96 & 0.06 & 0.078 & 0.006 & Cgh\\
  49731 & 2.379 & 0.202 & 5.25 & 15.1 & 5.22 & 1.10 & 0.065 & 0.066 & X\\
  49859 & 2.361 & 0.212 & 5.003 & 14.4 & 7.45 & 2.56 & 0.055 & 0.069 & Cgh\\
  50068 & 2.347 & 0.206 & 5.035 & 14.9 & - & - & - & - & L\\
  52870 & 2.359 & 0.205 & 4.938 & 15.2 & 5.53 & 0.05 & 0.053 & 0.009 & T\\
  52891 & 2.354 & 0.201 & 4.86 & 15.4 & 4.67 & 0.40 & 0.056 & 0.009 & B\\
  56349 & 2.352 & 0.209 & 5.314 & 15.2 & - & - & - & - & B\\
  65354 & 2.351 & 0.208 & 4.799 & 16.0 & 4.13 & 1.59 & 0.034 & 0.020 & Xc\\
  66309 & 2.353 & 0.208 & 4.873 & 15.3 & 3.45 & 0.46 & 0.094 & 0.023 & L\\
  66325 & 2.352 & 0.205 & 5.125 & 15.4 & 5.63 & 1.87 & 0.042 & 0.088 & Xc\\
  66403 & 2.387 & 0.202 & 5.004 & 15.4 & - & - & - & - & Ch\\
  67891 & 2.373 & 0.205 & 5.273 & 15.1 & 5.10 & 0.18 & 0.062 & 0.010 & Ch\\
  67918 & 2.376 & 0.203 & 4.942 & 15.1 & 4.76 & 1.27 & 0.076 & 0.069 & Ch\\
  67940 & 2.372 & 0.205 & 4.611 & 15.2 & 5.60 & 0.26 & 0.051 & 0.009 & Ch\\
  68114 & 2.36 & 0.209 & 4.635 & 15.5 & 5.96 & 1.15 & 0.031 & 0.016 & Cgh\\
  68685 & 2.358 & 0.218 & 5.458 & 15.5 & 4.66 & 0.13 & 0.056 & 0.011 & Xc\\
  69266 & 2.381 & 0.212 & 5.036 & 15.2 & 5.58 & 1.10 & 0.049 & 0.016 & X\\
  69706 & 2.343 & 0.2 & 4.808 & 15.3 & - & - & - & - & S\\
  70312 & 2.391 & 0.214 & 4.737 & 15.4 & 4.88 & 0.22 & 0.044 & 0.017 & B\\
  70361 & 2.357 & 0.206 & 4.453 & 15.0 & 6.49 & 1.22 & 0.043 & 0.018 & Ch\\
  70427 & 2.37 & 0.208 & 5.087 & 15.1 & 5.56 & 0.01 & 0.051 & 0.001 & X\\
  70511 & 2.402 & 0.203 & 5.165 & 15.2 & - & - & - & - & V\\
  71932 & 2.351 & 0.209 & 4.968 & 16.0 & 4.39 & 0.22 & 0.044 & 0.005 & Cgh\\
  72047 & 2.352 & 0.204 & 4.807 & 15.7 & 4.37 & 1.23 & 0.031 & 0.012 & Ch\\
  72143 & 2.354 & 0.209 & 5.232 & 15.8 & 4.71 & 0.12 & 0.046 & 0.006 & Cb\\
  72230 & 2.392 & 0.205 & 4.753 & 15.5 & 4.72 & 0.26 & 0.042 & 0.008 & B\\
  72292 & 2.358 & 0.212 & 5.534 & 15.9 & 4.57 & 0.86 & 0.025 & 0.014 & Xk\\
  72308 & 2.383 & 0.209 & 4.736 & 15.6 & 5.07 & 1.20 & 0.055 & 0.023 & Ch\\
  72384 & 2.381 & 0.209 & 5.281 & 15.5 & 3.95 & 0.64 & 0.088 & 0.049 & Ch\\
  72941 & 2.352 & 0.206 & 5.209 & 15.6 & 4.08 & 0.43 & 0.063 & 0.032 & X\\
  73860 & 2.378 & 0.209 & 5.093 & 16.1 & 4.32 & 1.09 & 0.042 & 0.026 & Cgh\\
  74755 & 2.39 & 0.201 & 5.303 & 15.7 & 3.92 & 0.03 & 0.072 & 0.006 & Cgh\\
  74962 & 2.359 & 0.213 & 5.034 & 15.9 & 4.36 & 0.32 & 0.041 & 0.015 & Ch\\
  75089 & 2.361 & 0.212 & 5.07 & 15.6 & 4.62 & 0.13 & 0.048 & 0.006 & Cgh\\
  76922 & 2.388 & 0.202 & 5.182 & 15.4 & - & - & - & - & Xk\\
  77421 & 2.374 & 0.206 & 4.815 & 15.6 & 4.52 & 0.09 & 0.041 & 0.004 & Xk\\
  78069 & 2.388 & 0.201 & 4.788 & 15.8 & 3.96 & 0.09 & 0.059 & 0.010 & Xk\\
  78826 & 2.356 & 0.217 & 5.52 & 15.7 & 4.93 & 0.97 & 0.042 & 0.032 & Xc\\
  78889 & 2.388 & 0.218 & 5.055 & 15.8 & 3.76 & 0.51 & 0.072 & 0.022 & T\\
  79044 & 2.386 & 0.209 & 4.739 & 16.0 & 3.77 & 0.52 & 0.063 & 0.020 & X\\
  85727 & 2.336 & 0.211 & 5.628 & 15.7 & - & - & - & - & S\\
  96405 & 2.342 & 0.214 & 5.347 & 15.4 & 5.11 & 0.35 & 0.056 & 0.018 & B\\
  96463 & 2.386 & 0.211 & 5.173 & 15.6 & 5.14 & 0.97 & 0.038 & 0.016 & Cb\\
  96768 & 2.377 & 0.206 & 4.686 & 15.5 & 4.57 & 0.85 & 0.059 & 0.025 & Cgh\\
  98345 & 2.361 & 0.211 & 5.137 & 15.4 & 4.61 & 0.25 & 0.063 & 0.015 & Cg\\
  100784 & 2.367 & 0.207 & 4.802 & 15.6 & 4.64 & 0.22 & 0.050 & 0.011 & B\\
  106794 & 2.339 & 0.215 & 5.317 & 15.8 & 3.82 & 0.61 & 0.070 & 0.078 & T\\
  107070 & 2.406 & 0.177 & 5.419 & 15.9 & - & - & - & - & L\\
  107742 & 2.399 & 0.199 & 5.48 & 16.1 & 3.39 & 0.64 & 0.067 & 0.022 & B\\
  111789 & 2.383 & 0.203 & 5.208 & 15.7 & 4.01 & 0.09 & 0.069 & 0.015 & Ch\\
  121096 & 2.346 & 0.211 & 5.338 & 15.8 & 4.10 & 0.15 & 0.036 & 0.005 & X\\
  129818 & 2.347 & 0.208 & 4.854 & 16.0 & 3.66 & 0.64 & 0.069 & 0.027 & Cb\\
  132056 & 2.35 & 0.205 & 5.17 & 15.9 & 3.96 & 0.47 & 0.053 & 0.015 & Ch\\
  132383 & 2.357 & 0.21 & 5.149 & 15.6 & 1.77 & 0.42 & 0.356 & 0.156 & L\\
  133123 & 2.361 & 0.214 & 5.261 & 16.2 & 3.24 & 0.04 & 0.062 & 0.009 & B\\
  133197 & 2.386 & 0.209 & 5.078 & 15.8 & 3.96 & 0.09 & 0.065 & 0.015 & T\\
  133503 & 2.349 & 0.211 & 5.042 & 16.3 & 3.02 & 0.02 & 0.058 & 0.008 & Cgh\\
  137397 & 2.393 & 0.202 & 4.972 & 16.3 & 2.83 & 0.43 & 0.067 & 0.027 & Ch\\
  162795 & 2.361 & 0.212 & 5.388 & 16.2 & 3.83 & 0.19 & 0.048 & 0.014 & T\\
  165536 & 2.354 & 0.21 & 5.064 & 16.0 & 3.58 & 0.65 & 0.057 & 0.030 & Cgh\\
  166264 & 2.381 & 0.177 & 5.199 & 15.9 & - & - & - & - & Xk\\
  169066 & 2.386 & 0.212 & 5.316 & 15.9 & 4.93 & 0.25 & 0.038 & 0.005 & X\\
  170184 & 2.347 & 0.208 & 5.234 & 16.3 & 3.69 & 0.85 & 0.039 & 0.014 & Xk\\
  174594 & 2.353 & 0.215 & 5.427 & 16.3 & 2.84 & 0.65 & 0.096 & 0.040 & Ch\\
  175811 & 2.379 & 0.214 & 4.802 & 15.7 & 3.91 & 0.57 & 0.057 & 0.027 & Cgh\\
  177258 & 2.325 & 0.214 & 5.162 & 16.9 & 2.49 & 0.57 & 0.050 & 0.028 & B\\
  178844 & 2.373 & 0.206 & 5.123 & 16.2 & 3.61 & 0.43 & 0.045 & 0.008 & Cgh\\
  186446 & 2.378 & 0.199 & 5.117 & 16.0 & - & - & - & - & Ch\\
  186714 & 2.404 & 0.215 & 5.207 & 16.9 & 2.33 & 0.53 & 0.062 & 0.028 & Cgh\\
  208048 & 2.389 & 0.213 & 4.844 & 16.5 & 3.26 & 0.38 & 0.052 & 0.013 & T\\
  210564 & 2.414 & 0.21 & 5.33 & 16.6 & 3.04 & 0.35 & 0.046 & 0.018 & X\\
  213825 & 2.356 & 0.207 & 4.882 & 16.1 & 3.40 & 0.60 & 0.061 & 0.021 & Cgh\\
  232922 & 2.348 & 0.215 & 5.17 & 16.0 & 3.78 & 0.15 & 0.054 & 0.009 & Ch\\
  242324 & 2.411 & 0.205 & 4.952 & 17.0 & 2.49 & 0.70 & 0.059 & 0.048 & Ch\\
  243648 & 2.387 & 0.193 & 5.093 & 16.9 & 2.56 & 0.05 & 0.052 & 0.008 & X\\
  250431 & 2.398 & 0.211 & 5.337 & 16.7 & 3.29 & 0.59 & 0.037 & 0.013 & T\\
  251796 & 2.341 & 0.204 & 4.826 & 16.8 & 3.08 & 0.91 & 0.035 & 0.022 & X\\
  252953 & 2.401 & 0.202 & 4.749 & 16.3 & 3.58 & 1.04 & 0.046 & 0.040 & Ch\\
  253538 & 2.393 & 0.217 & 5.526 & 16.4 & 3.17 & 0.91 & 0.048 & 0.044 & B\\
  253798 & 2.332 & 0.201 & 4.732 & 16.8 & 2.60 & 0.43 & 0.055 & 0.020 & Ch\\
  256789 & 2.373 & 0.207 & 4.857 & 16.8 & 2.54 & 0.32 & 0.052 & 0.010 & X\\
  262102 & 2.355 & 0.207 & 4.891 & 17.2 & 3.01 & 0.79 & 0.026 & 0.025 & Ch\\
  265259 & 2.404 & 0.21 & 4.981 & 16.9 & 3.18 & 0.58 & 0.036 & 0.015 & X\\
\end{longtable}
\end{longtab}


\begin{longtab}
\begin{longtable}{cccccccccc}
\caption{\label{table:results} Summary of the obtained results for the primitive asteroids in our sample.}\\
\hline\hline
Object & Class & Slope ($\%/1000\AA$) & Band & $\lambda_{cent} (\AA)$ & Depth ($\%$) \\
\hline
\endfirsthead
\caption{Continuation}\\
\hline\hline
Object & Class & Slope ($\%/1000\AA$) & Band & $\lambda_{cent} (\AA)$ & Depth ($\%$) \\
\hline
\endhead
\hline\\
\endfoot
  163 & Ch & 1.34$\pm$0.61 & YES & 6899$\pm$26 & 1.52$\pm$0.07\\
  10992 & Cgh & 0.95$\pm$0.63 & YES & 7074$\pm$37 & 1.80$\pm$0.14\\
  11856 & X & 4.94$\pm$0.71 & NO & - & -\\
  19415 & C & -0.56$\pm$0.66 & NO & - & -\\
  20992 & Xk & -0.17$\pm$0.64 & NO & - & -\\
  23397 & Ch & 1.95$\pm$0.65 & YES & 6957$\pm$78 & 4.43$\pm$0.22\\
  25381 & X & 2.16$\pm$0.63 & NO & - & -\\
  37437 & Xc & 3.65$\pm$0.76 & NO & - & -\\
  38173 & X & 4.90$\pm$0.70 & YES & 7149$\pm$60 & 4.32$\pm$0.28\\
  39694 & B & -0.77$\pm$0.65 & YES & 6887$\pm$98 & 2.23$\pm$0.30\\
  42155 & Cgh & 1.07$\pm$0.61 & YES & 6961$\pm$35 & 1.34$\pm$0.08\\
  42552 & Cgh & 1.33$\pm$0.62 & YES & 7107$\pm$33 & 2.26$\pm$0.10\\
  44766 & Cgh & 1.29$\pm$0.61 & YES & 7054$\pm$33 & 1.48$\pm$0.08\\
  44942 & X & 4.10$\pm$0.63 & YES & 7089$\pm$81 & 2.69$\pm$0.16\\
  45357 & Cgh & 1.75$\pm$0.62 & YES & 6913$\pm$68 & 2.76$\pm$0.11\\
  49731 & X & 2.85$\pm$0.62 & NO & - & -\\
  49859 & Cgh & 2.15$\pm$0.61 & YES & 6936$\pm$32 & 1.83$\pm$0.08\\
  52870 & T & 5.99$\pm$0.77 & NO & - & -\\
  52891 & B & -2.13$\pm$0.64 & NO & - & -\\
  56349 & B & -0.85$\pm$0.87 & NO & - & -\\
  65354 & Xc & 1.78$\pm$0.66 & NO & - & -\\
  66325 & Xc & 3.13$\pm$0.62 & NO & - & -\\
  66403 & Ch & 1.87$\pm$0.61 & YES & 7023$\pm$30 & 2.00$\pm$0.07\\
  67891 & Ch & 1.83$\pm$0.63 & YES & 7348$\pm$75 & 1.75$\pm$0.20\\
  67918 & Ch & 0.05$\pm$0.61 & YES & 7052$\pm$32 & 2.20$\pm$0.08\\
  67940 & Ch & 1.11$\pm$0.62 & YES & 7149$\pm$53 & 1.25$\pm$0.11\\
  68114 & Cgh & 1.63$\pm$0.61 & YES & 7153$\pm$33 & 1.29$\pm$0.07\\
  68685 & Xc & 2.18$\pm$0.74 & NO & - & -\\
  69266 & X & 3.07$\pm$0.61 & NO & - & -\\
  70312 & B & -0.74$\pm$0.62 & YES & 6996$\pm$35 & 2.00$\pm$0.10\\
  70361 & Ch & 0.63$\pm$0.61 & YES & 7049$\pm$32 & 1.96$\pm$0.08\\
  70427 & X & 2.65$\pm$0.61 & YES & 6890$\pm$68 & 1.39$\pm$0.14\\
  71932 & Cgh & 1.18$\pm$0.62 & YES & 7199$\pm$48 & 1.74$\pm$0.12\\
  72047 & Ch & 1.16$\pm$0.62 & YES & 7125$\pm$52 & 1.79$\pm$0.16\\
  72143 & Cb & 0.96$\pm$0.68 & NO & - & -\\
  72230 & B & -1.20$\pm$0.78 & NO & - & -\\
  72292 & Xk & 4.63$\pm$0.74 & NO & - & -\\
  72308 & Ch & 1.34$\pm$0.61 & YES & 7050$\pm$32 & 1.76$\pm$0.06\\
  72384 & Ch & 1.63$\pm$0.61 & YES & 7018$\pm$33 & 1.64$\pm$0.07\\
  72941 & X & 2.92$\pm$0.66 & YES & 7380$\pm$102 & 3.24$\pm$0.20\\
  73860 & Cgh & 1.81$\pm$0.61 & YES & 7230$\pm$65 & 2.49$\pm$0.12\\
  74755 & Cgh & 0.71$\pm$0.62 & YES & 7118$\pm$50 & 2.34$\pm$0.16\\
  74962 & Ch & 1.50$\pm$0.67 & YES & 6922$\pm$47 & 4.02$\pm$0.23\\
  75089 & Cgh & 1.65$\pm$0.68 & YES & 7200$\pm$108 & 4.15$\pm$0.37\\
  76922 & Xk & 5.06$\pm$0.78 & NO & - & -\\
  77421 & Xk & 1.78$\pm$0.65 & NO & - & -\\
  78069 & Xk & 2.51$\pm$0.76 & NO & - & -\\
  78826 & Xc & 2.22$\pm$0.61 & NO & - & -\\
  78889 & T & 3.62$\pm$0.66 & NO & - & -\\
  79044 & X & 1.76$\pm$0.63 & NO & - & -\\
  96405 & B & -2.02$\pm$0.63 & NO & - & -\\
  96463 & Cb & 2.49$\pm$0.65 & NO & - & -\\
  96768 & Cgh & 0.23$\pm$0.62 & YES & 6642$\pm$41 & 3.41$\pm$0.19\\
  98345 & Cg & -0.72$\pm$0.86 & NO & - & -\\
  100784 & B & -2.01$\pm$0.83 & NO & - & -\\
  106794 & T & 5.55$\pm$0.80 & YES & 7364$\pm$202 & 3.52$\pm$0.45\\
  107742 & B & -1.74$\pm$0.80 & NO & - & -\\
  111789 & Ch & 2.24$\pm$0.64 & YES & 7073$\pm$32 & 3.32$\pm$0.11\\
  121096 & X & 2.76$\pm$0.69 & YES & 6853$\pm$250 & 2.23$\pm$0.33\\
  129818 & Cb & 1.45$\pm$1.08 & NO & - & -\\
  132056 & Ch & 0.48$\pm$0.65 & YES & 7108$\pm$63 & 4.20$\pm$0.24\\
  133123 & B & -0.58$\pm$0.71 & YES & 7282$\pm$137 & 4.28$\pm$0.36\\
  133197 & T & 3.92$\pm$0.67 & NO & - & -\\
  133503 & Cgh & 1.15$\pm$0.63 & YES & 6898$\pm$76 & 2.83$\pm$0.27\\
  137397 & Ch & 0.92$\pm$0.62 & YES & 6740$\pm$46 & 2.67$\pm$0.14\\
  162795 & T & 2.93$\pm$0.65 & NO & - & -\\
  165536 & Cgh & 2.38$\pm$0.61 & YES & 7102$\pm$53 & 1.44$\pm$0.10\\
  166264 & Xk & 2.86$\pm$0.78 & NO & - & -\\
  169066 & X & 4.50$\pm$0.70 & YES & 6974$\pm$59 & 3.08$\pm$0.32\\
  170184 & Xk & 1.11$\pm$0.72 & NO & - & -\\
  174594 & Ch & 2.41$\pm$0.64 & YES & 7051$\pm$46 & 2.09$\pm$0.16\\
  175811 & Cgh & 0.61$\pm$0.61 & YES & 7093$\pm$31 & 2.36$\pm$0.07\\
  177258 & B & -1.41$\pm$0.81 & NO & - & -\\
  178844 & Cgh & 1.29$\pm$0.62 & YES & 7058$\pm$42 & 2.52$\pm$0.13\\
  186446 & Ch & 1.10$\pm$1.08 & YES & 7081$\pm$26 & 6.06$\pm$0.74\\
  186714 & Cgh & 3.25$\pm$0.68 & YES & 7145$\pm$66 & 5.50$\pm$0.29\\
  208048 & T & 5.74$\pm$0.64 & NO & - & -\\
  210564 & X & 7.22$\pm$0.78 & YES & 7009$\pm$24 & 9.14$\pm$0.58\\
  213825 & Cgh & 0.93$\pm$0.61 & YES & 7019$\pm$48 & 2.42$\pm$0.13\\
  232922 & Ch & -0.23$\pm$0.61 & YES & 7109$\pm$28 & 1.50$\pm$0.06\\
  242324 & Ch & 2.30$\pm$0.64 & YES & 7420$\pm$87 & 2.81$\pm$0.29\\
  243648 & X & 6.20$\pm$0.62 & YES & 6957$\pm$93 & 1.85$\pm$0.17\\
  250431 & T & 6.59$\pm$0.97 & NO & - & -\\
  251796 & X & 1.43$\pm$0.68 & NO & - & -\\
  252953 & Ch & 1.17$\pm$0.64 & YES & 7127$\pm$87 & 4.60$\pm$0.21\\
  253538 & B & -2.79$\pm$0.92 & YES & 7199$\pm$118 & 6.49$\pm$0.78\\
  253798 & Ch & 0.65$\pm$0.62 & YES & 6755$\pm$42 & 2.42$\pm$0.12\\
  256789 & X & 4.01$\pm$0.65 & NO & - & -\\
  262102 & Ch & 2.85$\pm$0.70 & YES & 6816$\pm$134 & 4.21$\pm$0.28\\
  265259 & X & 2.80$\pm$0.71 & NO & - & -\\
\end{longtable}
\end{longtab}

\bibliographystyle{aa} 
\bibliography{mybibliography.bib} 

\Online

\section*{Appendices}

\begin{appendix}
\section{Primitive spectra}
We present here the visible spectra of the primitive asteroids studied in this paper (a total of 90). The spectra are normalized to unity at 0.55 $\mu$m and binned (see Section \ref{reduction}). Visible spectra of asteroid (163) Erigone has been taken from the SMASS-II survey.

\begin{figure*}
\centering
\includegraphics[width=0.8\textwidth]{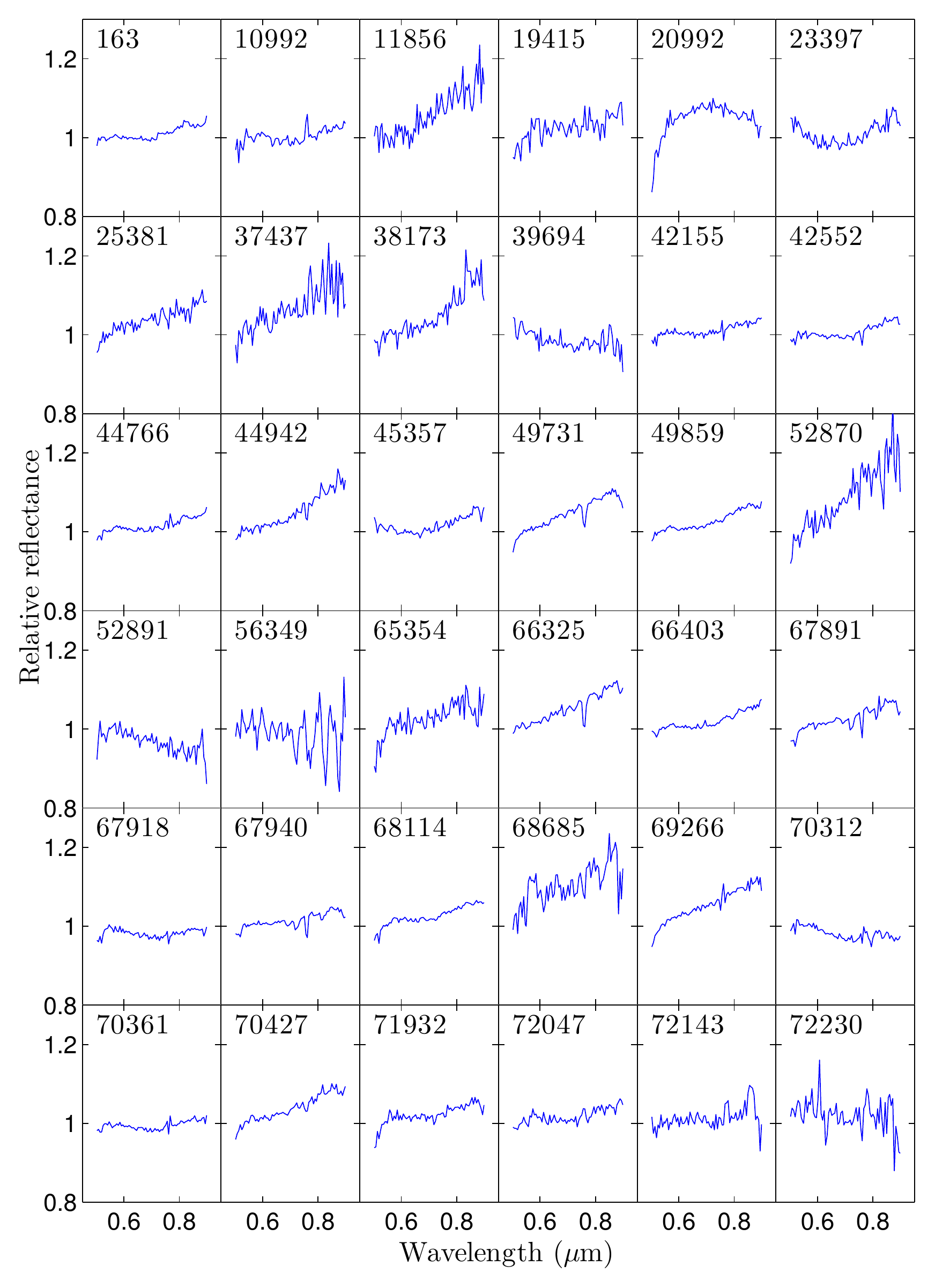}
\caption{Visible spectra of the primitive asteroids presented in this paper. Spectra are normalized to unity at 0.55 $\mu$m.}
\label{fig:final_spectra}
\end{figure*}

\addtocounter{figure}{-1}
\begin{figure*}
\centering
\includegraphics[width=0.8\textwidth]{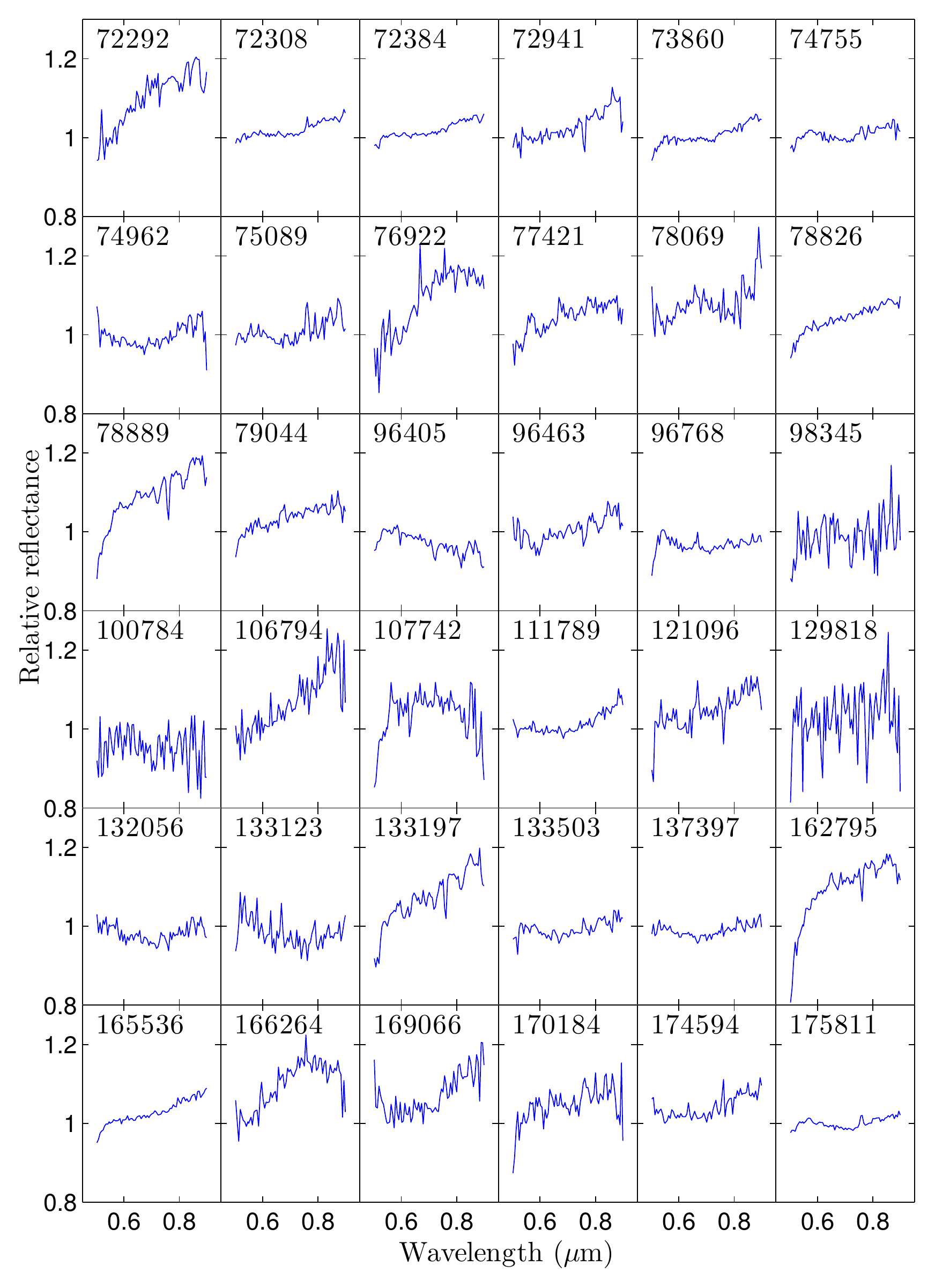}
\caption{Continued.}
\end{figure*}

\addtocounter{figure}{-1}
\begin{figure*}
\centering
\includegraphics[width=0.8\textwidth]{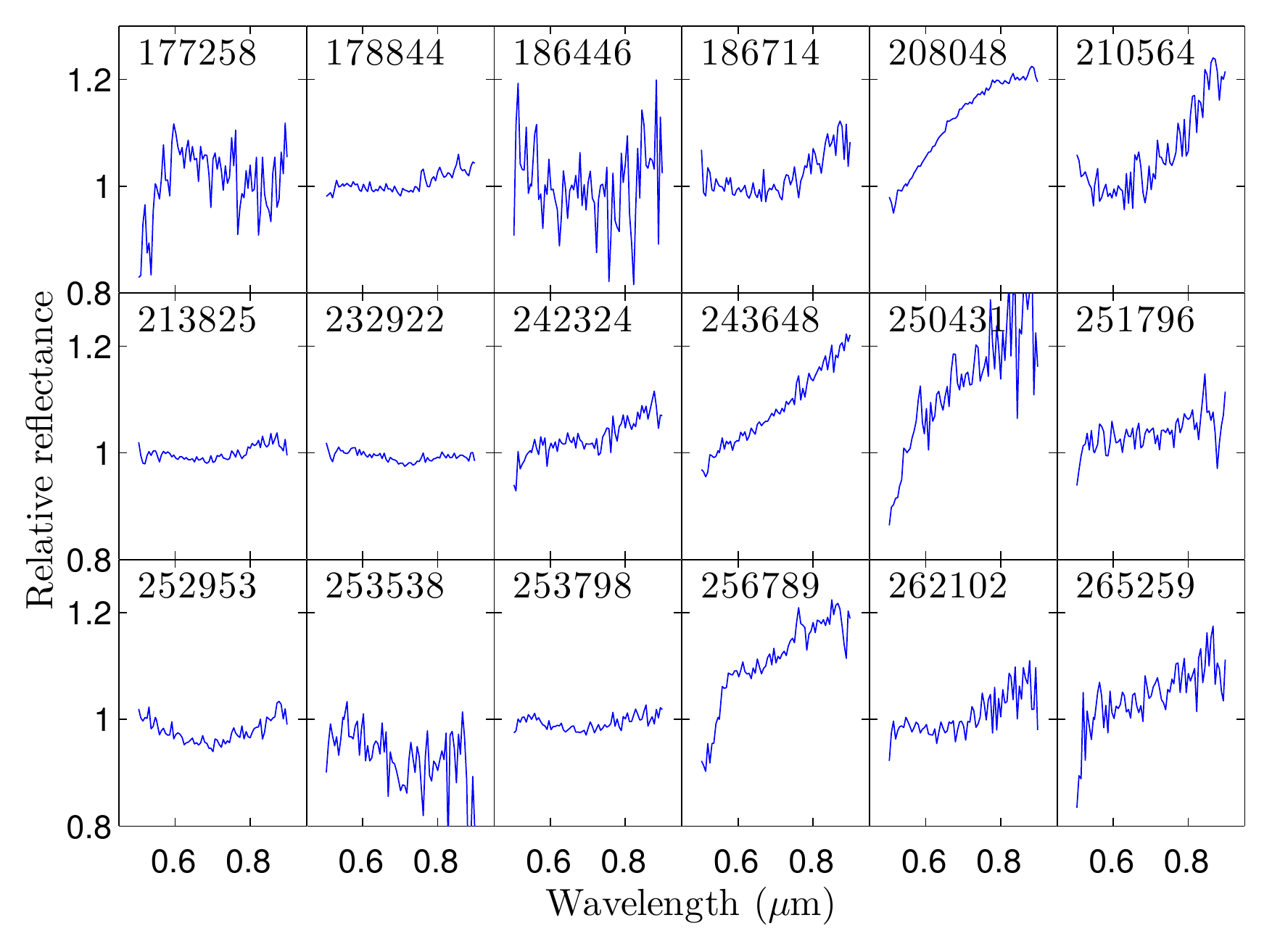}
\caption{Continued.}
\end{figure*}

\section{Non-primitive spectra}
We present here the visible spectra of the non-primitive asteroids studied in this paper (a total of 13). The spectra are normalized to unity at 0.55 $\mu$m and binned (see Section \ref{reduction}). Visible spectra of asteroid (571) Dulcinea has been taken from the SMASS-II survey.

\begin{figure*}
\centering
\includegraphics[width=0.8\textwidth]{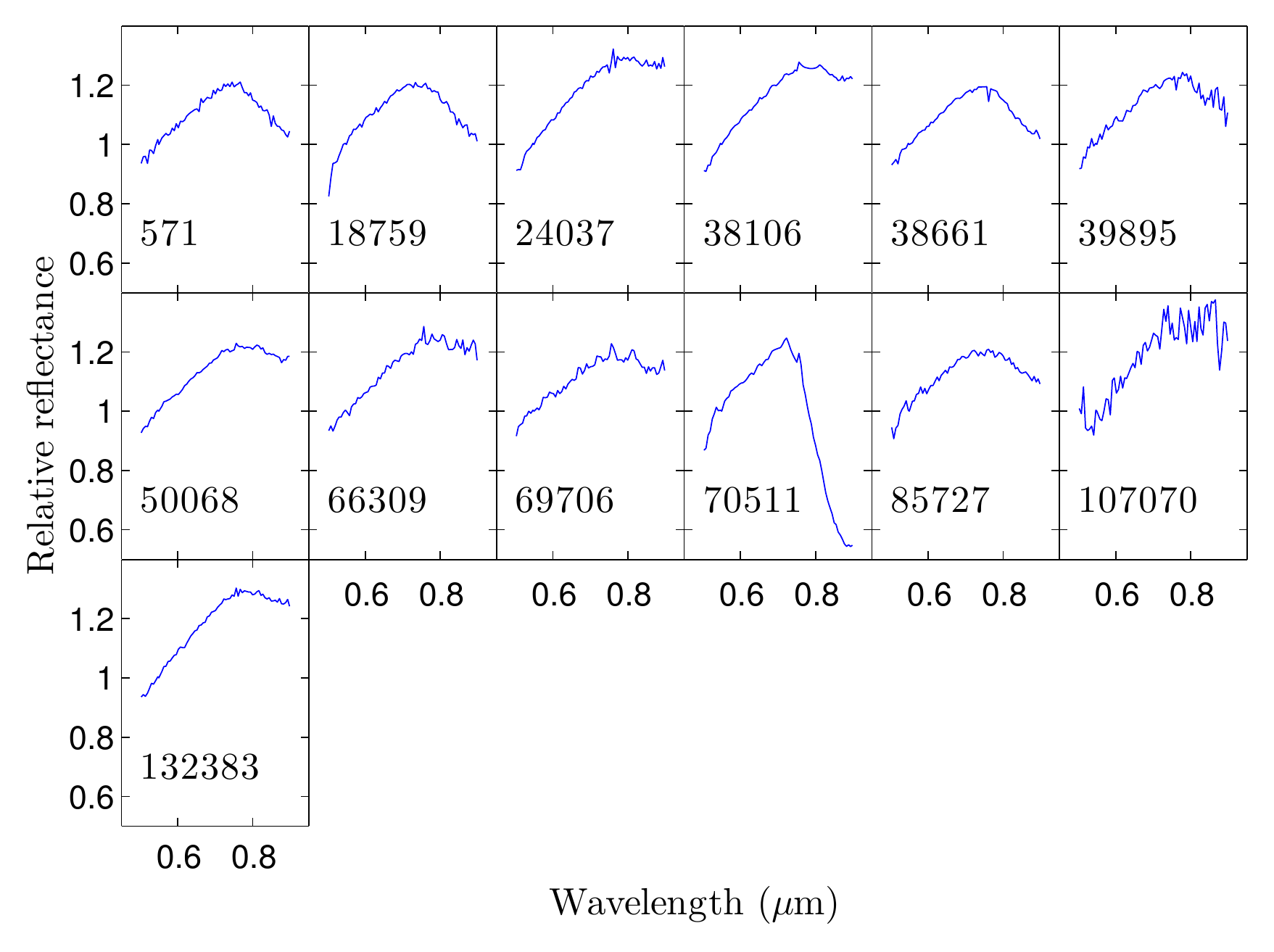}
\caption{Visible spectra of the non-primitive asteroids presented in this paper. Spectra are normalized to unity at 0.55 $\mu$m.}
\label{fig:final_spectra}
\end{figure*}

\end{appendix}

\end{document}